\documentclass[letterpaper, 10pt, conference]{ieeeconf}

\IEEEoverridecommandlockouts  
\usepackage[
  backend=biber,
  style=ieee,
  minnames=1,
  maxcitenames=2,
  maxbibnames=6,
  doi=false,
  isbn=false,
  url=false,
  natbib=true,
  clearlang=true,
  date=year,
]{biblatex}
\usepackage{xpatch}
\usepackage{xstring}

\DeclareSourcemap{
	\maps[datatype=bibtex]{
		\map[overwrite, foreach={booktitle,journaltitle,eventtitle,series,publisher,institution,type}]{
			\step[fieldsource=\regexp{$MAPLOOP}, match={XXX}, replace={Acad. Serbe Sci. Arts Glas Cl. Sci. Tech.}]
			\step[fieldsource=\regexp{$MAPLOOP}, match={XXX}, replace={Acta Acust.}]
			\step[fieldsource=\regexp{$MAPLOOP}, match={XXX}, replace={Acta Astron. (Poland)}]
			\step[fieldsource=\regexp{$MAPLOOP}, match={XXX}, replace={Acta Astron. Sin. (China)}]
			\step[fieldsource=\regexp{$MAPLOOP}, match={XXX}, replace={Acta Astronaut. (U.K.)}]
			\step[fieldsource=\regexp{$MAPLOOP}, match={XXX}, replace={Acta Astrophys. Sin. (China)}]
			\step[fieldsource=\regexp{$MAPLOOP}, match={XXX}, replace={Acta Autom. Sin.}]
			\step[fieldsource=\regexp{$MAPLOOP}, match={XXX}, replace={Acta Cienc. Indica Math.}]
			\step[fieldsource=\regexp{$MAPLOOP}, match={XXX}, replace={Acta Cienc. Indica Phys.}]
			\step[fieldsource=\regexp{$MAPLOOP}, match={XXX}, replace={Acta Crystallogr. A, Found. Crystallogr.}]
			\step[fieldsource=\regexp{$MAPLOOP}, match={XXX}, replace={Acta Crystallogr. B, Struct. Sci.}]
			\step[fieldsource=\regexp{$MAPLOOP}, match={XXX}, replace={Acta Crystallogr. C, Cryst. Struct. Commun.}]
			\step[fieldsource=\regexp{$MAPLOOP}, match={XXX}, replace={Acta Electron. (France)}]
			\step[fieldsource=\regexp{$MAPLOOP}, match={XXX}, replace={Acta Cyberno}]
			\step[fieldsource=\regexp{$MAPLOOP}, match={XXX}, replace={Acta Electron. Sin. (China)}]
			\step[fieldsource=\regexp{$MAPLOOP}, match={XXX}, replace={Acta Geod. Geophys. Montan. Hung.}]
			\step[fieldsource=\regexp{$MAPLOOP}, match={XXX}, replace={Acta Geophys. Pol.}]
			\step[fieldsource=\regexp{$MAPLOOP}, match={XXX}, replace={Acta Geophys. Sin. (China)}]
			\step[fieldsource=\regexp{$MAPLOOP}, match={XXX}, replace={Acta Geophys. Sin. (USA)}]
			\step[fieldsource=\regexp{$MAPLOOP}, match={XXX}, replace={Acta Metall.}]
			\step[fieldsource=\regexp{$MAPLOOP}, match={XXX}, replace={Acta Mex. Cienc. Tecnol.}]
			\step[fieldsource=\regexp{$MAPLOOP}, match={XXX}, replace={Acta Phys. Hung.}]
			\step[fieldsource=\regexp{$MAPLOOP}, match={XXX}, replace={Acta Phys. Pol. A}]
			\step[fieldsource=\regexp{$MAPLOOP}, match={XXX}, replace={Acta Phys. Pol. B}]
			\step[fieldsource=\regexp{$MAPLOOP}, match={XXX}, replace={Acta Phys. Sin.}]
			\step[fieldsource=\regexp{$MAPLOOP}, match={XXX}, replace={Acta Phys. Slovaca}]
			\step[fieldsource=\regexp{$MAPLOOP}, match={XXX}, replace={Acta Politec. Mex.}]
			\step[fieldsource=\regexp{$MAPLOOP}, match={XXX}, replace={Acta Polytech. Scand. Appl. Phys. Ser.}]
			\step[fieldsource=\regexp{$MAPLOOP}, match={XXX}, replace={Acta Polytech. Scand. Chem. Technol.}]
			\step[fieldsource=\regexp{$MAPLOOP}, match={XXX}, replace={Metall. Ser.}]
			\step[fieldsource=\regexp{$MAPLOOP}, match={XXX}, replace={Acta Polytech. Scand. Electr. Eng. Ser.}]
			\step[fieldsource=\regexp{$MAPLOOP}, match={XXX}, replace={Acta Polytech. Scand. Math. Comput. Sci. Ser.}]
			\step[fieldsource=\regexp{$MAPLOOP}, match={XXX}, replace={Acta Polytech. Scand. Mech. Eng. Ser.}]
			\step[fieldsource=\regexp{$MAPLOOP}, match={XXX}, replace={Acta Seismol. Sin.}]
			\step[fieldsource=\regexp{$MAPLOOP}, match={XXX}, replace={Acta Tech. Acad. Sci. Hung.}]
			\step[fieldsource=\regexp{$MAPLOOP}, match={XXX}, replace={Acta Tech. CSAV}]
			\step[fieldsource=\regexp{$MAPLOOP}, match={XXX}, replace={Acustica}]
			\step[fieldsource=\regexp{$MAPLOOP}, match={XXX}, replace={(AEU) Arch. Elektr. Ubertragung}]
			\step[fieldsource=\regexp{$MAPLOOP}, match={XXX}, replace={Akust. Zh.}]
			\step[fieldsource=\regexp{$MAPLOOP}, match={XXX}, replace={Algorithmica}]
			\step[fieldsource=\regexp{$MAPLOOP}, match={XXX}, replace={Alta Freq.}]
			\step[fieldsource=\regexp{$MAPLOOP}, match={XXX}, replace={An. Acad. Bras. Cienc.}]
			\step[fieldsource=\regexp{$MAPLOOP}, match={XXX}, replace={An. Fis.}]
			\step[fieldsource=\regexp{$MAPLOOP}, match={XXX}, replace={An. Mec. Electr.}]
			\step[fieldsource=\regexp{$MAPLOOP}, match={XXX}, replace={Angew. Inform.}]
			\step[fieldsource=\regexp{$MAPLOOP}, match={XXX}, replace={Ann. Inst. Henri Poincare Phys. Theor.}]
			\step[fieldsource=\regexp{$MAPLOOP}, match={XXX}, replace={Ann. Soc. Sci. Brux. I, Sci. Math. Astron. Phys.}]
			\step[fieldsource=\regexp{$MAPLOOP}, match={XXX}, replace={Arch. Elektr. Uebertrag. (AEU)}] 
			\step[fieldsource=\regexp{$MAPLOOP}, match={XXX}, replace={Arch. Elektron. Uebertrag. Tech.}] 
			\step[fieldsource=\regexp{$MAPLOOP}, match={XXX}, replace={Arch. Elektrotech. (Poland)}]
			\step[fieldsource=\regexp{$MAPLOOP}, match={XXX}, replace={Arch. Elektrotech. (Germany)}]
			\step[fieldsource=\regexp{$MAPLOOP}, match={XXX}, replace={Ark. Fys. Semin. Trondheim}]
			\step[fieldsource=\regexp{$MAPLOOP}, match={XXX}, replace={Astrofizika}]
			\step[fieldsource=\regexp{$MAPLOOP}, match={XXX}, replace={Astron. Nachr. (Germany)}]
			\step[fieldsource=\regexp{$MAPLOOP}, match={XXX}, replace={Astron. Tidsskr.}]
			\step[fieldsource=\regexp{$MAPLOOP}, match={XXX}, replace={Astron. Vestn.}]
			\step[fieldsource=\regexp{$MAPLOOP}, match={XXX}, replace={Astron. Zh.}]
			\step[fieldsource=\regexp{$MAPLOOP}, match={XXX}, replace={Atomwirtsch.-Atomtech.}]
			\step[fieldsource=\regexp{$MAPLOOP}, match={XXX}, replace={Atti Accad. Sci. Ist. Bologna CI. Sci. Fis.}]
			\step[fieldsource=\regexp{$MAPLOOP}, match={XXX}, replace={Rend. XIII}]
			\step[fieldsource=\regexp{$MAPLOOP}, match={XXX}, replace={Atti Accad. Sci. Torino I, CI. Sci. Fis. Math. Nat.}]
			\step[fieldsource=\regexp{$MAPLOOP}, match={XXX}, replace={Autom. Strum.}]
			\step[fieldsource=\regexp{$MAPLOOP}, match={XXX}, replace={Autom. Tech. Prax.}]
			\step[fieldsource=\regexp{$MAPLOOP}, match={XXX}, replace={Automatica}]
			\step[fieldsource=\regexp{$MAPLOOP}, match={XXX}, replace={Automatie}]
			\step[fieldsource=\regexp{$MAPLOOP}, match={XXX}, replace={Automatika}]
			\step[fieldsource=\regexp{$MAPLOOP}, match={XXX}, replace={Automatisierungstechnik}]
			\step[fieldsource=\regexp{$MAPLOOP}, match={XXX}, replace={Automatizace}]
			\step[fieldsource=\regexp{$MAPLOOP}, match={XXX}, replace={Automedica}]
			\step[fieldsource=\regexp{$MAPLOOP}, match={XXX}, replace={Avtom. Telemekh.}]
			\step[fieldsource=\regexp{$MAPLOOP}, match={XXX}, replace={Avtom. Vychisl. Tekh.}]
			\step[fieldsource=\regexp{$MAPLOOP}, match={XXX}, replace={Avtomatika}]
			\step[fieldsource=\regexp{$MAPLOOP}, match={XXX}, replace={Avtometriya}]
			\step[fieldsource=\regexp{$MAPLOOP}, match={ATZelektronik worldwide}, replace={ATZelektron. worldw.}]
			\step[fieldsource=\regexp{$MAPLOOP}, match={XXX}, replace={Ber. Bunsenges. Phys. Chem.}]
			\step[fieldsource=\regexp{$MAPLOOP}, match={XXX}, replace={Biofizika}]
			\step[fieldsource=\regexp{$MAPLOOP}, match={XXX}, replace={Biometrika}]
			\step[fieldsource=\regexp{$MAPLOOP}, match={XXX}, replace={Boll. Geofis. Teor. Appl.}]
			\step[fieldsource=\regexp{$MAPLOOP}, match={XXX}, replace={Bull. Acad. Serbe Sci. Arts cl. Sci. Tech.}]
			\step[fieldsource=\regexp{$MAPLOOP}, match={XXX}, replace={Bull. Annu. Soc. Suisse Chronom. Lab. Suisse.}]
			\step[fieldsource=\regexp{$MAPLOOP}, match={XXX}, replace={Rech. Horlog.}]
			\step[fieldsource=\regexp{$MAPLOOP}, match={XXX}, replace={Bull. Cl. Sci. Acad. R. Belg.}]
			\step[fieldsource=\regexp{$MAPLOOP}, match={XXX}, replace={Bull. Dir. Etud. Rech. A}]
			\step[fieldsource=\regexp{$MAPLOOP}, match={XXX}, replace={Bull. Dir. Etud. Rech. B}]
			\step[fieldsource=\regexp{$MAPLOOP}, match={XXX}, replace={Bull. Dir. Etud. Rech. C}]
			\step[fieldsource=\regexp{$MAPLOOP}, match={XXX}, replace={Bull. Liaison Rech. Inform. Autom.}]
			\step[fieldsource=\regexp{$MAPLOOP}, match={XXX}, replace={Bur. Etud. Autom.}]
			\step[fieldsource=\regexp{$MAPLOOP}, match={XXX}, replace={CFI-Ceram. Forum Int.-Ber. Dtsch. Keram. Ges.}]
			\step[fieldsource=\regexp{$MAPLOOP}, match={XXX}, replace={Chem. Scr.}]
			\step[fieldsource=\regexp{$MAPLOOP}, match={XXX}, replace={Ciel Terre}]
			\step[fieldsource=\regexp{$MAPLOOP}, match={XXX}, replace={Cybernetica}]
			\step[fieldsource=\regexp{$MAPLOOP}, match={XXX}, replace={Deut. Hydrogr. Z.}]
			\step[fieldsource=\regexp{$MAPLOOP}, match={XXX}, replace={Dokl. Akad. Nauk SSSR}]
			\step[fieldsource=\regexp{$MAPLOOP}, match={XXX}, replace={Electroacoustique}]
			\step[fieldsource=\regexp{$MAPLOOP}, match={XXX}, replace={Electrochim. Acta}]
			\step[fieldsource=\regexp{$MAPLOOP}, match={XXX}, replace={Electrochim. Metal.}]
			\step[fieldsource=\regexp{$MAPLOOP}, match={XXX}, replace={Elektor Electron.}]
			\step[fieldsource=\regexp{$MAPLOOP}, match={XXX}, replace={Elektr. Bahnen}]
			\step[fieldsource=\regexp{$MAPLOOP}, match={XXX}, replace={Elektr. Energ.-Tech.}]
			\step[fieldsource=\regexp{$MAPLOOP}, match={XXX}, replace={Elektr. Masch.}]
			\step[fieldsource=\regexp{$MAPLOOP}, match={XXX}, replace={Elektr. Stn.}]
			\step[fieldsource=\regexp{$MAPLOOP}, match={XXX}, replace={Elektrichestvo}]
			\step[fieldsource=\regexp{$MAPLOOP}, match={XXX}, replace={Elektrie}]
			\step[fieldsource=\regexp{$MAPLOOP}, match={XXX}, replace={Elektrizitaetswirtschaft}]
			\step[fieldsource=\regexp{$MAPLOOP}, match={XXX}, replace={Elektro}]
			\step[fieldsource=\regexp{$MAPLOOP}, match={XXX}, replace={Elektro-Anz.}]
			\step[fieldsource=\regexp{$MAPLOOP}, match={XXX}, replace={Elektro-Jahr}]
			\step[fieldsource=\regexp{$MAPLOOP}, match={XXX}, replace={Elektrokhimiya}]
			\step[fieldsource=\regexp{$MAPLOOP}, match={XXX}, replace={Elektron}]
			\step[fieldsource=\regexp{$MAPLOOP}, match={XXX}, replace={Elektron Int.}]
			\step[fieldsource=\regexp{$MAPLOOP}, match={XXX}, replace={Elektron. Entwitkl.}]
			\step[fieldsource=\regexp{$MAPLOOP}, match={XXX}, replace={Elektron. Ind}]
			\step[fieldsource=\regexp{$MAPLOOP}, match={XXX}, replace={Elektron. J.}]
			\step[fieldsource=\regexp{$MAPLOOP}, match={XXX}, replace={Elektron. Prax.}]
			\step[fieldsource=\regexp{$MAPLOOP}, match={XXX}, replace={Elektron. Tekh.}]
			\step[fieldsource=\regexp{$MAPLOOP}, match={XXX}, replace={Elektronica}]
			\step[fieldsource=\regexp{$MAPLOOP}, match={XXX}, replace={Elektronik}]
			\step[fieldsource=\regexp{$MAPLOOP}, match={XXX}, replace={Elektronika}]
			\step[fieldsource=\regexp{$MAPLOOP}, match={XXX}, replace={Elektroniker}]
			\step[fieldsource=\regexp{$MAPLOOP}, match={XXX}, replace={Elektronikschau}]
			\step[fieldsource=\regexp{$MAPLOOP}, match={XXX}, replace={Elektrosvyaz}]
			\step[fieldsource=\regexp{$MAPLOOP}, match={XXX}, replace={Elektrotech. Cas.}]
			\step[fieldsource=\regexp{$MAPLOOP}, match={Elektrotechnik und Informationstechnik}, replace={Elektrotech. Inf. Tech.}]
			\step[fieldsource=\regexp{$MAPLOOP}, match={XXX}, replace={Elektrotech. Obz.}]
			\step[fieldsource=\regexp{$MAPLOOP}, match={XXX}, replace={Elektrotechniek}]
			\step[fieldsource=\regexp{$MAPLOOP}, match={XXX}, replace={Elektrotechnik (Czechoslovakia)}]
			\step[fieldsource=\regexp{$MAPLOOP}, match={XXX}, replace={Elektrotechnik (Switzerland)}]
			\step[fieldsource=\regexp{$MAPLOOP}, match={XXX}, replace={Elektrotechnik (Germany)}]
			\step[fieldsource=\regexp{$MAPLOOP}, match={XXX}, replace={Elektrotechnika}]
			\step[fieldsource=\regexp{$MAPLOOP}, match={XXX}, replace={Elektrotehnika, Zagreb}]
			\step[fieldsource=\regexp{$MAPLOOP}, match={XXX}, replace={Elektrotekhnika}]
			\step[fieldsource=\regexp{$MAPLOOP}, match={XXX}, replace={Elektroteknikeren}]
			\step[fieldsource=\regexp{$MAPLOOP}, match={XXX}, replace={Elektrowaerme Int. B.}]
			\step[fieldsource=\regexp{$MAPLOOP}, match={XXX}, replace={Elettrificazione}]
			\step[fieldsource=\regexp{$MAPLOOP}, match={XXX}, replace={Elettron. Oggi}]
			\step[fieldsource=\regexp{$MAPLOOP}, match={XXX}, replace={Elettron. Telecomun.}]
			\step[fieldsource=\regexp{$MAPLOOP}, match={XXX}, replace={Elettrotecnica}]
			\step[fieldsource=\regexp{$MAPLOOP}, match={XXX}, replace={Elek. Med Aktuell Elektron.}]
			\step[fieldsource=\regexp{$MAPLOOP}, match={XXX}, replace={Elteknik}]
			\step[fieldsource=\regexp{$MAPLOOP}, match={XXX}, replace={Energ. Atomtech.}]
			\step[fieldsource=\regexp{$MAPLOOP}, match={XXX}, replace={Energ. Elettr.}]
			\step[fieldsource=\regexp{$MAPLOOP}, match={XXX}, replace={Energetica}]
			\step[fieldsource=\regexp{$MAPLOOP}, match={XXX}, replace={Energetik}]
			\step[fieldsource=\regexp{$MAPLOOP}, match={XXX}, replace={Energetika}]
			\step[fieldsource=\regexp{$MAPLOOP}, match={XXX}, replace={Energetyka}]
			\step[fieldsource=\regexp{$MAPLOOP}, match={XXX}, replace={Energia Nuclear}]
			\step[fieldsource=\regexp{$MAPLOOP}, match={XXX}, replace={Energie Technik (Germany)}]
			\step[fieldsource=\regexp{$MAPLOOP}, match={XXX}, replace={Energie Technik (Switzerland)}]
			\step[fieldsource=\regexp{$MAPLOOP}, match={XXX}, replace={Entropie}]
			\step[fieldsource=\regexp{$MAPLOOP}, match={XXX}, replace={ETZ}]
			\step[fieldsource=\regexp{$MAPLOOP}, match={XXX}, replace={ETZ Arch.}]
			\step[fieldsource=\regexp{$MAPLOOP}, match={XXX}, replace={Feingeraetetechnik}]
			\step[fieldsource=\regexp{$MAPLOOP}, match={XXX}, replace={Feinw. Tech. Messtech.}]
			\step[fieldsource=\regexp{$MAPLOOP}, match={XXX}, replace={Fert. Tech. Betr.}]
			\step[fieldsource=\regexp{$MAPLOOP}, match={XXX}, replace={Fis. Tecnol.}]
			\step[fieldsource=\regexp{$MAPLOOP}, match={XXX}, replace={Fiz. Khim. Obrab. Mater.}]
			\step[fieldsource=\regexp{$MAPLOOP}, match={XXX}, replace={Fiz. Met. Metalloved}]
			\step[fieldsource=\regexp{$MAPLOOP}, match={XXX}, replace={Fiz. Nizk. Temp.}]
			\step[fieldsource=\regexp{$MAPLOOP}, match={XXX}, replace={Fiz. Plazmy}]
			\step[fieldsource=\regexp{$MAPLOOP}, match={XXX}, replace={Fiz. Tekh. Poluprovodn.}]
			\step[fieldsource=\regexp{$MAPLOOP}, match={XXX}, replace={Fiz. Tverd. Tela}]
			\step[fieldsource=\regexp{$MAPLOOP}, match={XXX}, replace={Fiz.-Khim. Mekh. Mater.}]
			\step[fieldsource=\regexp{$MAPLOOP}, match={XXX}, replace={Fizika}]
			\step[fieldsource=\regexp{$MAPLOOP}, match={XXX}, replace={Forsch.-Ber. Landes Nordrh.-Westfal.}]
			\step[fieldsource=\regexp{$MAPLOOP}, match={XXX}, replace={Frequenz}]
			\step[fieldsource=\regexp{$MAPLOOP}, match={XXX}, replace={Fys. Tidsskr.}]
			\step[fieldsource=\regexp{$MAPLOOP}, match={XXX}, replace={G. Fis.}]
			\step[fieldsource=\regexp{$MAPLOOP}, match={XXX}, replace={Geliotekhnika}]
			\step[fieldsource=\regexp{$MAPLOOP}, match={XXX}, replace={Geochim. Cosmochim. Acta}]
			\step[fieldsource=\regexp{$MAPLOOP}, match={XXX}, replace={Haerterei-Tech. Mitt.}]
			\step[fieldsource=\regexp{$MAPLOOP}, match={XXX}, replace={Helv. Chim. Acta}]
			\step[fieldsource=\regexp{$MAPLOOP}, match={XXX}, replace={Helv. Med. Acta}]
			\step[fieldsource=\regexp{$MAPLOOP}, match={XXX}, replace={Helv. Phys. Acta}]
			\step[fieldsource=\regexp{$MAPLOOP}, match={XXX}, replace={Hochfreq. Electroakust}]
			\step[fieldsource=\regexp{$MAPLOOP}, match={XXX}, replace={Hoppe-Seylers Z. Physiol. Chem.}]
			\step[fieldsource=\regexp{$MAPLOOP}, match={XXX}, replace={Inf. Elektron.}]
			\step[fieldsource=\regexp{$MAPLOOP}, match={XXX}, replace={Inf. Elettron.}]
			\step[fieldsource=\regexp{$MAPLOOP}, match={XXX}, replace={Inform. Forsch. Entwickl.}]
			\step[fieldsource=\regexp{$MAPLOOP}, match={XXX}, replace={Inform. Spektrum}]
			\step[fieldsource=\regexp{$MAPLOOP}, match={XXX}, replace={Inform.-Fachber.}]
			\step[fieldsource=\regexp{$MAPLOOP}, match={XXX}, replace={Informatie}]
			\step[fieldsource=\regexp{$MAPLOOP}, match={XXX}, replace={Informatik}]
			\step[fieldsource=\regexp{$MAPLOOP}, match={XXX}, replace={Informatologia Yugosl.}]
			\step[fieldsource=\regexp{$MAPLOOP}, match={XXX}, replace={Informatyka}]
			\step[fieldsource=\regexp{$MAPLOOP}, match={XXX}, replace={Infowelt}]
			\step[fieldsource=\regexp{$MAPLOOP}, match={XXX}, replace={Ing. Electr. Mec.}]
			\step[fieldsource=\regexp{$MAPLOOP}, match={XXX}, replace={Ing. Mec. Electr.}]
			\step[fieldsource=\regexp{$MAPLOOP}, match={XXX}, replace={Ing.-Arch.}]
			\step[fieldsource=\regexp{$MAPLOOP}, match={XXX}, replace={Inzh.-Fiz. Zh.}]
			\step[fieldsource=\regexp{$MAPLOOP}, match={XXX}, replace={Izmer. Tekh.}]
			\step[fieldsource=\regexp{$MAPLOOP}, match={XXX}, replace={Izv. Akad. Nauk Arm. SSR Ser. Tekh. Nauk}]
			\step[fieldsource=\regexp{$MAPLOOP}, match={XXX}, replace={Izv. Akad. Nauk SSSR Energ. Transp.}]
			\step[fieldsource=\regexp{$MAPLOOP}, match={XXX}, replace={Izv. Akad. Nauk SSSR Fiz. Atmos. Okeana}]
			\step[fieldsource=\regexp{$MAPLOOP}, match={XXX}, replace={Izv. Akad. Nauk SSSR Fiz. Zemli}]
			\step[fieldsource=\regexp{$MAPLOOP}, match={XXX}, replace={Izv. Akad. Nauk SSSR Ser. Fiz.}]
			\step[fieldsource=\regexp{$MAPLOOP}, match={XXX}, replace={Izv. Vyssh. Uchebn. Zaved. Elektromekh.}]
			\step[fieldsource=\regexp{$MAPLOOP}, match={XXX}, replace={Izv. Vyssh. Uchebn. Zaved. Radioelektron.}]
			\step[fieldsource=\regexp{$MAPLOOP}, match={XXX}, replace={Izv. Vyssh. Uchebn. Zaved. Radiofiz.}]
			\step[fieldsource=\regexp{$MAPLOOP}, match={XXX}, replace={J. Chim Phys. Phys.-Chim Biol.}]
			\step[fieldsource=\regexp{$MAPLOOP}, match={XXX}, replace={Kernenergie}]
			\step[fieldsource=\regexp{$MAPLOOP}, match={XXX}, replace={Kerntechnik}]
			\step[fieldsource=\regexp{$MAPLOOP}, match={XXX}, replace={Khim. Fiz.}]
			\step[fieldsource=\regexp{$MAPLOOP}, match={XXX}, replace={Kibern. Vychisl. Tekh.}]
			\step[fieldsource=\regexp{$MAPLOOP}, match={XXX}, replace={Kibernetika}]
			\step[fieldsource=\regexp{$MAPLOOP}, match={XXX}, replace={Kristallografiya}]
			\step[fieldsource=\regexp{$MAPLOOP}, match={XXX}, replace={Kvantovaya Elektron. Mosk.}]
			\step[fieldsource=\regexp{$MAPLOOP}, match={XXX}, replace={Kybernetes}]
			\step[fieldsource=\regexp{$MAPLOOP}, match={XXX}, replace={Kybernetika}]
			\step[fieldsource=\regexp{$MAPLOOP}, match={XXX}, replace={Med. Tek.}]
			\step[fieldsource=\regexp{$MAPLOOP}, match={XXX}, replace={Mekh. Avtom. Proizvod.}]
			\step[fieldsource=\regexp{$MAPLOOP}, match={XXX}, replace={Meres Autom.}]
			\step[fieldsource=\regexp{$MAPLOOP}, match={XXX}, replace={Mesures}]
			\step[fieldsource=\regexp{$MAPLOOP}, match={XXX}, replace={Metallofizika}]
			\step[fieldsource=\regexp{$MAPLOOP}, match={XXX}, replace={Metalloved. Term. Obrab. Met.}]
			\step[fieldsource=\regexp{$MAPLOOP}, match={XXX}, replace={Meteorol. Gidrol.}]
			\step[fieldsource=\regexp{$MAPLOOP}, match={XXX}, replace={Meteorol. Rundsch.}]
			\step[fieldsource=\regexp{$MAPLOOP}, match={XXX}, replace={Metrol. Apl.}]
			\step[fieldsource=\regexp{$MAPLOOP}, match={XXX}, replace={Metrologia}]
			\step[fieldsource=\regexp{$MAPLOOP}, match={XXX}, replace={Medel. Simul.}]
			\step[fieldsource=\regexp{$MAPLOOP}, match={XXX}, replace={Nachr. Dok.}]
			\step[fieldsource=\regexp{$MAPLOOP}, match={XXX}, replace={Nachr.tech. Elektron.}]
			\step[fieldsource=\regexp{$MAPLOOP}, match={XXX}, replace={Naturwissentchaften}]
			\step[fieldsource=\regexp{$MAPLOOP}, match={XXX}, replace={Neue Tech.}]
			\step[fieldsource=\regexp{$MAPLOOP}, match={XXX}, replace={Neue Tech. Buero}]
			\step[fieldsource=\regexp{$MAPLOOP}, match={XXX}, replace={Nukleonika}]
			\step[fieldsource=\regexp{$MAPLOOP}, match={XXX}, replace={Numer. Math.}]
			\step[fieldsource=\regexp{$MAPLOOP}, match={XXX}, replace={Nuovo Cimento A}]
			\step[fieldsource=\regexp{$MAPLOOP}, match={XXX}, replace={Nuovo Cimento B}]
			\step[fieldsource=\regexp{$MAPLOOP}, match={XXX}, replace={Nouvo Cimento C}]
			\step[fieldsource=\regexp{$MAPLOOP}, match={XXX}, replace={Nuovo Cimento D}]
			\step[fieldsource=\regexp{$MAPLOOP}, match={XXX}, replace={Okeanologiya}]
			\step[fieldsource=\regexp{$MAPLOOP}, match={XXX}, replace={Opt. Spektrosk.}]
			\step[fieldsource=\regexp{$MAPLOOP}, match={XXX}, replace={Opt.-Mekh. Prom.}]
			\step[fieldsource=\regexp{$MAPLOOP}, match={XXX}, replace={Optik}]
			\step[fieldsource=\regexp{$MAPLOOP}, match={XXX}, replace={Photogrammetria}]
			\step[fieldsource=\regexp{$MAPLOOP}, match={XXX}, replace={Photonics Spectra}]
			\step[fieldsource=\regexp{$MAPLOOP}, match={XXX}, replace={Pis’ma Astron. Zh. }]
			\step[fieldsource=\regexp{$MAPLOOP}, match={XXX}, replace={Pis’ma Zh. Eksp. Teor. Fiz. }]
			\step[fieldsource=\regexp{$MAPLOOP}, match={XXX}, replace={Pis’ma Zh. Tekh. Fiz. }]
			\step[fieldsource=\regexp{$MAPLOOP}, match={XXX}, replace={Poverkhn., Fiz. Khim. Mekh.}]
			\step[fieldsource=\regexp{$MAPLOOP}, match={XXX}, replace={Pr. Inst. Elektrotech.}]
			\step[fieldsource=\regexp{$MAPLOOP}, match={XXX}, replace={Prib. Sist. Upr.}]
			\step[fieldsource=\regexp{$MAPLOOP}, match={XXX}, replace={Prib. Tekh. Eksp.}]
			\step[fieldsource=\regexp{$MAPLOOP}, match={XXX}, replace={Prikl. Mat. Mekh.}]
			\step[fieldsource=\regexp{$MAPLOOP}, match={XXX}, replace={Prikl. Mekh.}]
			\step[fieldsource=\regexp{$MAPLOOP}, match={XXX}, replace={Probl. Kibern.}]
			\step[fieldsource=\regexp{$MAPLOOP}, match={XXX}, replace={Probl. Peredachi Inf.}]
			\step[fieldsource=\regexp{$MAPLOOP}, match={XXX}, replace={Proc. K. Ned. Akad. Wet. B, Palaeontol.}]
			\step[fieldsource=\regexp{$MAPLOOP}, match={XXX}, replace={Anthropol. }]
			\step[fieldsource=\regexp{$MAPLOOP}, match={XXX}, replace={Programmirovanie}]
			\step[fieldsource=\regexp{$MAPLOOP}, match={XXX}, replace={Prz. Elektrotech.}]
			\step[fieldsource=\regexp{$MAPLOOP}, match={XXX}, replace={Prz. Telekomun.}]
			\step[fieldsource=\regexp{$MAPLOOP}, match={XXX}, replace={PT/Elektrotech. Elektron.}]
			\step[fieldsource=\regexp{$MAPLOOP}, match={XXX}, replace={Radio Fernsehen Elektron.}]
			\step[fieldsource=\regexp{$MAPLOOP}, match={XXX}, replace={Radiotekh. Elektron.}]
			\step[fieldsource=\regexp{$MAPLOOP}, match={XXX}, replace={Radiotekhnika Mosk.}]
			\step[fieldsource=\regexp{$MAPLOOP}, match={XXX}, replace={Rev. Acad. Cienc. Zaragoza}]
			\step[fieldsource=\regexp{$MAPLOOP}, match={XXX}, replace={Rev. Electrotec. (Argentina)}]
			\step[fieldsource=\regexp{$MAPLOOP}, match={XXX}, replace={Rev. Electrotec. (Spain)}]
			\step[fieldsource=\regexp{$MAPLOOP}, match={XXX}, replace={Rev. Energ.}]
			\step[fieldsource=\regexp{$MAPLOOP}, match={XXX}, replace={Rev. Esp. Electron. Rev. Geofis.}]
			\step[fieldsource=\regexp{$MAPLOOP}, match={XXX}, replace={Ric. Autom.}]
			\step[fieldsource=\regexp{$MAPLOOP}, match={XXX}, replace={Ric. Spettrosc.}]
			\step[fieldsource=\regexp{$MAPLOOP}, match={XXX}, replace={Robotersysteme}]
			\step[fieldsource=\regexp{$MAPLOOP}, match={XXX}, replace={Rozpr. Electrotech.}]
			\step[fieldsource=\regexp{$MAPLOOP}, match={XXX}, replace={Sadhana}]
			\step[fieldsource=\regexp{$MAPLOOP}, match={XXX}, replace={Schweiz. Tech. Z.}]
			\step[fieldsource=\regexp{$MAPLOOP}, match={XXX}, replace={Scientia}]
			\step[fieldsource=\regexp{$MAPLOOP}, match={XXX}, replace={Siemens Forsch. Entwickl. Ber.}]
			\step[fieldsource=\regexp{$MAPLOOP}, match={XXX}, replace={Sist. Autom.}]
			\step[fieldsource=\regexp{$MAPLOOP}, match={XXX}, replace={Sitzungsber. Oester. Akad. Wiss. Math.-}]
			\step[fieldsource=\regexp{$MAPLOOP}, match={XXX}, replace={Naturwiss. Kl. Abt. II (Austria)}]
			\step[fieldsource=\regexp{$MAPLOOP}, match={XXX}, replace={Spectrochim. Acta A, Mol. Spectrosc.}]
			\step[fieldsource=\regexp{$MAPLOOP}, match={XXX}, replace={Spectrochim. Acta B, At. Spectrosc.}]
			\step[fieldsource=\regexp{$MAPLOOP}, match={XXX}, replace={Sprache Datenverarb.}]
			\step[fieldsource=\regexp{$MAPLOOP}, match={XXX}, replace={Stanki Instrum.}]
			\step[fieldsource=\regexp{$MAPLOOP}, match={XXX}, replace={Steklo Keram.}]
			\step[fieldsource=\regexp{$MAPLOOP}, match={XXX}, replace={Svetotekhnika  TE Int.}]
			\step[fieldsource=\regexp{$MAPLOOP}, match={XXX}, replace={Tech. Bull. Vevey}]
			\step[fieldsource=\regexp{$MAPLOOP}, match={XXX}, replace={Tech. Mitt. Krupp (Engl. Ed.)}]
			\step[fieldsource=\regexp{$MAPLOOP}, match={XXX}, replace={Tech. Mitt. PTT}]
			\step[fieldsource=\regexp{$MAPLOOP}, match={XXX}, replace={Tech. Mitt. RFZ}]
			\step[fieldsource=\regexp{$MAPLOOP}, match={XXX}, replace={Technica}]
			\step[fieldsource=\regexp{$MAPLOOP}, match={XXX}, replace={Tecnica}]
			\step[fieldsource=\regexp{$MAPLOOP}, match={XXX}, replace={Teh. Fiz.}]
			\step[fieldsource=\regexp{$MAPLOOP}, match={XXX}, replace={Tehnika}]
			\step[fieldsource=\regexp{$MAPLOOP}, match={XXX}, replace={Tekh. Elektrodin.}]
			\step[fieldsource=\regexp{$MAPLOOP}, match={XXX}, replace={Tekh. Kibern.}]
			\step[fieldsource=\regexp{$MAPLOOP}, match={XXX}, replace={Tekh. Kino Telev.}]
			\step[fieldsource=\regexp{$MAPLOOP}, match={XXX}, replace={Tekh. Misul}]
			\step[fieldsource=\regexp{$MAPLOOP}, match={XXX}, replace={Telekomunikacije}]
			\step[fieldsource=\regexp{$MAPLOOP}, match={XXX}, replace={Telektronikk}]
			\step[fieldsource=\regexp{$MAPLOOP}, match={XXX}, replace={Teleteknik}]
			\step[fieldsource=\regexp{$MAPLOOP}, match={XXX}, replace={Teor. Mat. Fiz.}]
			\step[fieldsource=\regexp{$MAPLOOP}, match={XXX}, replace={Teploeoergetika}]
			\step[fieldsource=\regexp{$MAPLOOP}, match={XXX}, replace={Teplofiz. Vvs. Temp.}]
			\step[fieldsource=\regexp{$MAPLOOP}, match={XXX}, replace={Tidskr. Dok.}]
			\step[fieldsource=\regexp{$MAPLOOP}, match={XXX}, replace={TN Nachr.}]
			\step[fieldsource=\regexp{$MAPLOOP}, match={XXX}, replace={Toute Electron.}]
			\step[fieldsource=\regexp{$MAPLOOP}, match={XXX}, replace={Tr. Inst. Teor. Astron.}]
			\step[fieldsource=\regexp{$MAPLOOP}, match={XXX}, replace={Ukr. Fiz. Zh.}]
			\step[fieldsource=\regexp{$MAPLOOP}, match={XXX}, replace={Usp. Fiz. Nauk}]
			\step[fieldsource=\regexp{$MAPLOOP}, match={XXX}, replace={Vak.-Tech.}]
			\step[fieldsource=\regexp{$MAPLOOP}, match={XXX}, replace={VDE Fachiber.}]
			\step[fieldsource=\regexp{$MAPLOOP}, match={XXX}, replace={VDI Z.}]
			\step[fieldsource=\regexp{$MAPLOOP}, match={XXX}, replace={Vestn. Mashinostr.}]
			\step[fieldsource=\regexp{$MAPLOOP}, match={XXX}, replace={Vestn. Mosk. Univ. 15, Vychisl. Mat. Kibern.}]
			\step[fieldsource=\regexp{$MAPLOOP}, match={XXX}, replace={Vestn. Mosk. Univ. 3, Fiz. Astron.}]
			\step[fieldsource=\regexp{$MAPLOOP}, match={XXX}, replace={Vesti Akad. Navuk BSSR Ser. Fiz. Energ. Navuk}]
			\step[fieldsource=\regexp{$MAPLOOP}, match={XXX}, replace={VGB Kraftwerkstech. (Ger. Ed.)}]
			\step[fieldsource=\regexp{$MAPLOOP}, match={XXX}, replace={Vide Couches Minces}]
			\step[fieldsource=\regexp{$MAPLOOP}, match={XXX}, replace={Vistas Astron.}]
			\step[fieldsource=\regexp{$MAPLOOP}, match={XXX}, replace={Vopr. At. Nauki Tekh. Ser., Fiz. Radiats.}]
			\step[fieldsource=\regexp{$MAPLOOP}, match={XXX}, replace={Povrezhdenii Radiats. Materialoved.}]
			\step[fieldsource=\regexp{$MAPLOOP}, match={XXX}, replace={Vopr. At. Nauki Tekh. Ser., Obshch. Yad. Fiz.}]
			\step[fieldsource=\regexp{$MAPLOOP}, match={XXX}, replace={Vuoto Sci. Tecnol.}]
			\step[fieldsource=\regexp{$MAPLOOP}, match={XXX}, replace={Wiss. Z. Friedrich-Schiller-Univ. Jena}]
			\step[fieldsource=\regexp{$MAPLOOP}, match={XXX}, replace={Nat.wiss. Reihe}]
			\step[fieldsource=\regexp{$MAPLOOP}, match={XXX}, replace={Wiss. Z. Karl-Marx-Univ. Leipz. Math.-}]
			\step[fieldsource=\regexp{$MAPLOOP}, match={XXX}, replace={Nat.wiss. Reihe}]
			\step[fieldsource=\regexp{$MAPLOOP}, match={XXX}, replace={Wiss. Z. Tech. Hochsch. Ilmenau}]
			\step[fieldsource=\regexp{$MAPLOOP}, match={XXX}, replace={Wiss. Z. Tech. Univ. Dresd.}]
			\step[fieldsource=\regexp{$MAPLOOP}, match={XXX}, replace={Wiss. Z. Tech. Univ. Karl-Marx-Stadt}]
			\step[fieldsource=\regexp{$MAPLOOP}, match={XXX}, replace={Yad. Fiz.}]
			\step[fieldsource=\regexp{$MAPLOOP}, match={XXX}, replace={Z. Angew. Math. Mech.}]
			\step[fieldsource=\regexp{$MAPLOOP}, match={XXX}, replace={Z. Angew. Math. Phys.}]
			\step[fieldsource=\regexp{$MAPLOOP}, match={XXX}, replace={Z. Met.kd.}]
			\step[fieldsource=\regexp{$MAPLOOP}, match={XXX}, replace={Z. Nat. Forsch. A, Phys. Phys. Chem. Kosmophys.}]
			\step[fieldsource=\regexp{$MAPLOOP}, match={XXX}, replace={Z. Oper. Res. A, Theor.}]
			\step[fieldsource=\regexp{$MAPLOOP}, match={XXX}, replace={Z. Oper. Res. B, Prax.}]
			\step[fieldsource=\regexp{$MAPLOOP}, match={XXX}, replace={Z. Phys.}]
			\step[fieldsource=\regexp{$MAPLOOP}, match={XXX}, replace={Z. Phys. A, At. Nuclei}]
			\step[fieldsource=\regexp{$MAPLOOP}, match={XXX}, replace={Z. Phys. B, Condens. Matter}]
			\step[fieldsource=\regexp{$MAPLOOP}, match={XXX}, replace={Z. Phys. C, Part Fields}]
			\step[fieldsource=\regexp{$MAPLOOP}, match={XXX}, replace={Z. Phys. Chem. Neue Folge}]
			\step[fieldsource=\regexp{$MAPLOOP}, match={XXX}, replace={Z. Phys. Chem., Leipz.}]
			\step[fieldsource=\regexp{$MAPLOOP}, match={XXX}, replace={Z. Phys. D, At. Mol. Clusters}]
			\step[fieldsource=\regexp{$MAPLOOP}, match={XXX}, replace={Zavod. Lab.}]
			\step[fieldsource=\regexp{$MAPLOOP}, match={XXX}, replace={Zh. Eksp. Teor. Fiz.}]
			\step[fieldsource=\regexp{$MAPLOOP}, match={XXX}, replace={Zh. Fiz. Khim.}]
			\step[fieldsource=\regexp{$MAPLOOP}, match={XXX}, replace={Zh. Prikl. Mekh. Tekh. Fiz.}]
			\step[fieldsource=\regexp{$MAPLOOP}, match={XXX}, replace={Zh. Prikl. Spektrosk.}]
			\step[fieldsource=\regexp{$MAPLOOP}, match={XXX}, replace={Zh. Tekh. Fiz.}]
			\step[fieldsource=\regexp{$MAPLOOP}, match={XXX}, replace={Zh. Vychisl. Mat. Mat. Fiz.}]
			\step[fieldsource=\regexp{$MAPLOOP}, match={XXX}, replace={Zisin, J. Seismol. Soc. Jpn.}] 
			\step[fieldsource=\regexp{$MAPLOOP}, match={Production}, replace={Prod.}]
			\step[fieldsource=\regexp{$MAPLOOP}, match={Reliability}, replace={Rel.}]
			\step[fieldsource=\regexp{$MAPLOOP}, match={Report}, replace={Rep.}]
			\step[fieldsource=\regexp{$MAPLOOP}, match={Semiconductor}, replace={Semicond.}]
			\step[fieldsource=\regexp{$MAPLOOP}, match={Research}, replace={Res.}]
			\step[fieldsource=\regexp{$MAPLOOP}, match={Sensing}, replace={Sens.}]
			\step[fieldsource=\regexp{$MAPLOOP}, match={Resonance}, replace={Reson.}]
			\step[fieldsource=\regexp{$MAPLOOP}, match={Series}, replace={Ser.}]
			\step[fieldsource=\regexp{$MAPLOOP}, match={Resources}, replace={Resour.}]
			\step[fieldsource=\regexp{$MAPLOOP}, match={Simulation}, replace={Simul.}]
			\step[fieldsource=\regexp{$MAPLOOP}, match={Reviews}, replace={Rev.}]
			\step[fieldsource=\regexp{$MAPLOOP}, match={Review}, replace={Rev.}]
			\step[fieldsource=\regexp{$MAPLOOP}, match={Singapore}, replace={Singap.}]
			\step[fieldsource=\regexp{$MAPLOOP}, match={Robotics}, replace={Robot.}]
			\step[fieldsource=\regexp{$MAPLOOP}, match={Sistema}, replace={Sist.}]
			\step[fieldsource=\regexp{$MAPLOOP}, match={Royal}, replace={Roy.}]
			\step[fieldsource=\regexp{$MAPLOOP}, match={Society}, replace={Soc.}]
			\step[fieldsource=\regexp{$MAPLOOP}, match={Safety}, replace={Saf.}]
			\step[fieldsource=\regexp{$MAPLOOP}, match={Sociological}, replace={Sociol.}]
			\step[fieldsource=\regexp{$MAPLOOP}, match={Satellite}, replace={Satell.}]
			\step[fieldsource=\regexp{$MAPLOOP}, match={Software}, replace={Softw.}]
			\step[fieldsource=\regexp{$MAPLOOP}, match={Scandinavian}, replace={Scand.}]
			\step[fieldsource=\regexp{$MAPLOOP}, match={Solar}, replace={Sol.}]
			\step[fieldsource=\regexp{$MAPLOOP}, match={Sciences}, replace={Sci.}]
			\step[fieldsource=\regexp{$MAPLOOP}, match={Science}, replace={Sci.}]
			\step[fieldsource=\regexp{$MAPLOOP}, match={Soviet}, replace={Sov.}]
			\step[fieldsource=\regexp{$MAPLOOP}, match={Section}, replace={Sect.}]
			\step[fieldsource=\regexp{$MAPLOOP}, match={Spectroscopy}, replace={Spectrosc.}]
			\step[fieldsource=\regexp{$MAPLOOP}, match={Security}, replace={Secur.}]
			\step[fieldsource=\regexp{$MAPLOOP}, match={Spectrum}, replace={Spectr.}]
			\step[fieldsource=\regexp{$MAPLOOP}, match={Seismology}, replace={Seismol.}]
			\step[fieldsource=\regexp{$MAPLOOP}, match={Speculations}, replace={Specul.}]
			\step[fieldsource=\regexp{$MAPLOOP}, match={Selected}, replace={Sel.}]
			\step[fieldsource=\regexp{$MAPLOOP}, match={Statistics}, replace={Statist.}]
			\step[fieldsource=\regexp{$MAPLOOP}, match={Structures}, replace={Struct.}]
			\step[fieldsource=\regexp{$MAPLOOP}, match={Structure}, replace={Struct.}]
			\step[fieldsource=\regexp{$MAPLOOP}, match={Terrestrial}, replace={Terr.}]
			\step[fieldsource=\regexp{$MAPLOOP}, match={Studies}, replace={Stud.}]
			\step[fieldsource=\regexp{$MAPLOOP}, match={Theoretical}, replace={Theor.}]
			\step[fieldsource=\regexp{$MAPLOOP}, match={Superconductivity}, replace={Supercond.}]
			\step[fieldsource=\regexp{$MAPLOOP}, match={Transactions}, replace={Trans.}]
			\step[fieldsource=\regexp{$MAPLOOP}, match={Supplement}, replace={Suppl.}]
			\step[fieldsource=\regexp{$MAPLOOP}, match={Translation}, replace={Transl.}]
			\step[fieldsource=\regexp{$MAPLOOP}, match={Surface}, replace={Surf.}]
			\step[fieldsource=\regexp{$MAPLOOP}, match={Transmission}, replace={Transmiss.}]
			\step[fieldsource=\regexp{$MAPLOOP}, match={Survey}, replace={Surv.}]
			\step[fieldsource=\regexp{$MAPLOOP}, match={Transportation}, replace={Transp.}]
			\step[fieldsource=\regexp{$MAPLOOP}, match={Sustainable}, replace={Sustain.}]
			\step[fieldsource=\regexp{$MAPLOOP}, match={Tutorials}, replace={Tut.}]
			\step[fieldsource=\regexp{$MAPLOOP}, match={Symposium}, replace={Symp.}]
			\step[fieldsource=\regexp{$MAPLOOP}, match={Ultrasonic}, replace={Ultrason.}]
			\step[fieldsource=\regexp{$MAPLOOP}, match={Systems}, replace={Syst.}]
			\step[fieldsource=\regexp{$MAPLOOP}, match={System}, replace={Syst.}]
			\step[fieldsource=\regexp{$MAPLOOP}, match={University}, replace={Univ.}]
			\step[fieldsource=\regexp{$MAPLOOP}, match={\detokenize{Universität}}, replace={Univ.}]
			\step[fieldsource=\regexp{$MAPLOOP}, match={\detokenize{Université}}, replace={Univ.}]
			\step[fieldsource=\regexp{$MAPLOOP}, match={Technical}, replace={Tech.}]
			\step[fieldsource=\regexp{$MAPLOOP}, match={Technische}, replace={Tech.}]
			\step[fieldsource=\regexp{$MAPLOOP}, match={Vacuum}, replace={Vac.}]
			\step[fieldsource=\regexp{$MAPLOOP}, match={Techniques}, replace={Techn.}]
			\step[fieldsource=\regexp{$MAPLOOP}, match={Vehicles}, replace={Veh.}]
			\step[fieldsource=\regexp{$MAPLOOP}, match={Vehicle}, replace={Veh.}]
			\step[fieldsource=\regexp{$MAPLOOP}, match={Vehicular}, replace={Veh.}]
			\step[fieldsource=\regexp{$MAPLOOP}, match={Technology}, replace={Technol.}]
			\step[fieldsource=\regexp{$MAPLOOP}, match={Technological}, replace={Technol.}]
			\step[fieldsource=\regexp{$MAPLOOP}, match={Vibration}, replace={Vib.}]
			\step[fieldsource=\regexp{$MAPLOOP}, match={Telecommunications}, replace={Telecommun.}]
			\step[fieldsource=\regexp{$MAPLOOP}, match={Visual}, replace={Vis.}]
			\step[fieldsource=\regexp{$MAPLOOP}, match={Television}, replace={Telev.}]
			\step[fieldsource=\regexp{$MAPLOOP}, match={Welding}, replace={Weld.}]
			\step[fieldsource=\regexp{$MAPLOOP}, match={Temperature}, replace={Temp.}]
			\step[fieldsource=\regexp{$MAPLOOP}, match={Working}, replace={Work.}]
			\step[fieldsource=\regexp{$MAPLOOP}, match={Learning}, replace={Learn.}]
			\step[fieldsource=\regexp{$MAPLOOP}, match={Measurement}, replace={Meas.}]
			\step[fieldsource=\regexp{$MAPLOOP}, match={Letters}, replace={Lett.}]
			\step[fieldsource=\regexp{$MAPLOOP}, match={Letter}, replace={Lett.}]
			\step[fieldsource=\regexp{$MAPLOOP}, match={Mechanical}, replace={Mech.}]
			\step[fieldsource=\regexp{$MAPLOOP}, match={Mechanics}, replace={Mech.}]
			\step[fieldsource=\regexp{$MAPLOOP}, match={Mechanic}, replace={Mech.}]
			\step[fieldsource=\regexp{$MAPLOOP}, match={Lightwave}, replace={Lightw.}]
			\step[fieldsource=\regexp{$MAPLOOP}, match={Medical}, replace={Med.}]
			\step[fieldsource=\regexp{$MAPLOOP}, match={Logic, Logical}, replace={Log.}]
			\step[fieldsource=\regexp{$MAPLOOP}, match={Metals}, replace={Met.}]
			\step[fieldsource=\regexp{$MAPLOOP}, match={Luminescence}, replace={Lumin.}]
			\step[fieldsource=\regexp{$MAPLOOP}, match={Metallurgy}, replace={Metall.}]
			\step[fieldsource=\regexp{$MAPLOOP}, match={Machines}, replace={Mach.}]
			\step[fieldsource=\regexp{$MAPLOOP}, match={Machine}, replace={Mach.}]
			\step[fieldsource=\regexp{$MAPLOOP}, match={Meteorology}, replace={Meteorol.}]
			\step[fieldsource=\regexp{$MAPLOOP}, match={Magazine}, replace={Mag.}]
			\step[fieldsource=\regexp{$MAPLOOP}, match={Metropolitan}, replace={Metrop.}]
			\step[fieldsource=\regexp{$MAPLOOP}, match={Magnetics}, replace={Magn.}]
			\step[fieldsource=\regexp{$MAPLOOP}, match={Mexican, Mexico}, replace={Mex.}]
			\step[fieldsource=\regexp{$MAPLOOP}, match={Management}, replace={Manage.}]
			\step[fieldsource=\regexp{$MAPLOOP}, match={Microelectromechanical}, replace={Microelectromech.}]
			\step[fieldsource=\regexp{$MAPLOOP}, match={Managing}, replace={Manag.}]
			\step[fieldsource=\regexp{$MAPLOOP}, match={Microgravity}, replace={Microgr.}]
			\step[fieldsource=\regexp{$MAPLOOP}, match={Manufacturing}, replace={Manuf.}]
			\step[fieldsource=\regexp{$MAPLOOP}, match={Microscopy}, replace={Microsc.}]
			\step[fieldsource=\regexp{$MAPLOOP}, match={Marine}, replace={Mar.}]
			\step[fieldsource=\regexp{$MAPLOOP}, match={Microwaves}, replace={Microw.}]
			\step[fieldsource=\regexp{$MAPLOOP}, match={Microwave}, replace={Microw.}]
			\step[fieldsource=\regexp{$MAPLOOP}, match={Materials}, replace={Mater.}]
			\step[fieldsource=\regexp{$MAPLOOP}, match={Material}, replace={Mater.}]
			\step[fieldsource=\regexp{$MAPLOOP}, match={Military}, replace={Mil.}]
			\step[fieldsource=\regexp{$MAPLOOP}, match={Modeling}, replace={Model.}]
			\step[fieldsource=\regexp{$MAPLOOP}, match={Modelling}, replace={Model.}]
			\step[fieldsource=\regexp{$MAPLOOP}, match={Oceanic}, replace={Ocean.}]
			\step[fieldsource=\regexp{$MAPLOOP}, match={Molecular}, replace={Mol.}]
			\step[fieldsource=\regexp{$MAPLOOP}, match={Oceanography}, replace={Oceanogr.}]
			\step[fieldsource=\regexp{$MAPLOOP}, match={Monitoring}, replace={Monit.}]
			\step[fieldsource=\regexp{$MAPLOOP}, match={Occupation}, replace={Occupat.}]
			\step[fieldsource=\regexp{$MAPLOOP}, match={Multiphysics}, replace={Multiphys.}]
			\step[fieldsource=\regexp{$MAPLOOP}, match={Operational}, replace={Oper.}]
			\step[fieldsource=\regexp{$MAPLOOP}, match={Nanobioscience}, replace={Nanobiosci.}]
			\step[fieldsource=\regexp{$MAPLOOP}, match={Optical}, replace={Opt.}]
			\step[fieldsource=\regexp{$MAPLOOP}, match={Nanotechnology}, replace={Nanotechnol.}]
			\step[fieldsource=\regexp{$MAPLOOP}, match={Optics}, replace={Opt.}]
			\step[fieldsource=\regexp{$MAPLOOP}, match={National}, replace={Nat.}]
			\step[fieldsource=\regexp{$MAPLOOP}, match={Optimization}, replace={Optim.}]
			\step[fieldsource=\regexp{$MAPLOOP}, match={Naval}, replace={Nav.}]
			\step[fieldsource=\regexp{$MAPLOOP}, match={Organization}, replace={Org.}]
			\step[fieldsource=\regexp{$MAPLOOP}, match={Networking}, replace={Netw.}]
			\step[fieldsource=\regexp{$MAPLOOP}, match={Networked}, replace={Netw.}]
			\step[fieldsource=\regexp{$MAPLOOP}, match={Network}, replace={Netw.}]
			\step[fieldsource=\regexp{$MAPLOOP}, match={Packaging}, replace={Packag.}]
			\step[fieldsource=\regexp{$MAPLOOP}, match={Newsletter}, replace={Newslett.}]
			\step[fieldsource=\regexp{$MAPLOOP}, match={Particle}, replace={Part.}]
			\step[fieldsource=\regexp{$MAPLOOP}, match={Nondestructive}, replace={Nondestruct.}]
			\step[fieldsource=\regexp{$MAPLOOP}, match={Patent}, replace={Pat.}]
			\step[fieldsource=\regexp{$MAPLOOP}, match={Nuclear}, replace={Nucl.}]
			\step[fieldsource=\regexp{$MAPLOOP}, match={Performance}, replace={Perform.}]
			\step[fieldsource=\regexp{$MAPLOOP}, match={Numerical}, replace={Numer.}]
			\step[fieldsource=\regexp{$MAPLOOP}, match={Personal}, replace={Pers.}]
			\step[fieldsource=\regexp{$MAPLOOP}, match={Observations}, replace={Observ.}]
			\step[fieldsource=\regexp{$MAPLOOP}, match={Philosophical}, replace={Philos.}]
			\step[fieldsource=\regexp{$MAPLOOP}, match={Photonics}, replace={Photon.}]
			\step[fieldsource=\regexp{$MAPLOOP}, match={Productivity}, replace={Productiv.}]
			\step[fieldsource=\regexp{$MAPLOOP}, match={Photovoltaics}, replace={Photovolt.}]
			\step[fieldsource=\regexp{$MAPLOOP}, match={Programming}, replace={Program.}]
			\step[fieldsource=\regexp{$MAPLOOP}, match={Physics}, replace={Phys.}]
			\step[fieldsource=\regexp{$MAPLOOP}, match={Progress}, replace={Prog.}]
			\step[fieldsource=\regexp{$MAPLOOP}, match={Physiology}, replace={Physiol.}]
			\step[fieldsource=\regexp{$MAPLOOP}, match={Propagation}, replace={Propag.}]
			\step[fieldsource=\regexp{$MAPLOOP}, match={Planetary}, replace={Planet.}]
			\step[fieldsource=\regexp{$MAPLOOP}, match={Psychology}, replace={Psychol.}]
			\step[fieldsource=\regexp{$MAPLOOP}, match={Pneumatics}, replace={Pneum.}]
			\step[fieldsource=\regexp{$MAPLOOP}, match={Quality}, replace={Qual.}]
			\step[fieldsource=\regexp{$MAPLOOP}, match={Pollution}, replace={Pollut.}]
			\step[fieldsource=\regexp{$MAPLOOP}, match={Quarterly}, replace={Quart.}]
			\step[fieldsource=\regexp{$MAPLOOP}, match={Polymer}, replace={Polym.}]
			\step[fieldsource=\regexp{$MAPLOOP}, match={Radiation}, replace={Radiat.}]
			\step[fieldsource=\regexp{$MAPLOOP}, match={Polytechnic}, replace={Polytech.}]
			\step[fieldsource=\regexp{$MAPLOOP}, match={Radiology}, replace={Radiol.}]
			\step[fieldsource=\regexp{$MAPLOOP}, match={Practice}, replace={Pract.}]
			\step[fieldsource=\regexp{$MAPLOOP}, match={Reactor}, replace={React.}]
			\step[fieldsource=\regexp{$MAPLOOP}, match={Precision}, replace={Precis.}]
			\step[fieldsource=\regexp{$MAPLOOP}, match={Receivers}, replace={Receiv.}]
			\step[fieldsource=\regexp{$MAPLOOP}, match={Principles}, replace={Princ.}]
			\step[fieldsource=\regexp{$MAPLOOP}, match={Recognition}, replace={Recognit.}]
			\step[fieldsource=\regexp{$MAPLOOP}, match={Proceedings}, replace={Proc.}]
			\step[fieldsource=\regexp{$MAPLOOP}, match={Record}, replace={Rec.}]
			\step[fieldsource=\regexp{$MAPLOOP}, match={Processing}, replace={Process.}]
			\step[fieldsource=\regexp{$MAPLOOP}, match={Rehabilitation}, replace={Rehabil.}]
			\step[fieldsource=\regexp{$MAPLOOP}, match={Conversion}, replace={Convers.}]
			\step[fieldsource=\regexp{$MAPLOOP}, match={Digital}, replace={Digit.}]
			\step[fieldsource=\regexp{$MAPLOOP}, match={Convention}, replace={Conv.}]
			\step[fieldsource=\regexp{$MAPLOOP}, match={Disclosure}, replace={Discl.}]
			\step[fieldsource=\regexp{$MAPLOOP}, match={Correspondence}, replace={Corresp.}]
			\step[fieldsource=\regexp{$MAPLOOP}, match={Discussions}, replace={Discuss.}]
			\step[fieldsource=\regexp{$MAPLOOP}, match={Critical}, replace={Crit.}]
			\step[fieldsource=\regexp{$MAPLOOP}, match={Dissertations}, replace={Diss.}]
			\step[fieldsource=\regexp{$MAPLOOP}, match={Crystal}, replace={Cryst.}]
			\step[fieldsource=\regexp{$MAPLOOP}, match={Distributed}, replace={Distrib.}]
			\step[fieldsource=\regexp{$MAPLOOP}, match={Crystallography}, replace={Crystallogr.}]
			\step[fieldsource=\regexp{$MAPLOOP}, match={Dynamics}, replace={Dyn.}]
			\step[fieldsource=\regexp{$MAPLOOP}, match={Cybernetics}, replace={Cybern.}]
			\step[fieldsource=\regexp{$MAPLOOP}, match={Earthquake}, replace={Earthq.}]
			\step[fieldsource=\regexp{$MAPLOOP}, match={Decision}, replace={Decis.}]
			\step[fieldsource=\regexp{$MAPLOOP}, match={Economics}, replace={Econ.}]
			\step[fieldsource=\regexp{$MAPLOOP}, match={Economic}, replace={Econ.}]
			\step[fieldsource=\regexp{$MAPLOOP}, match={Economical}, replace={Econ.}]
			\step[fieldsource=\regexp{$MAPLOOP}, match={Edition}, replace={Ed.}]
			\step[fieldsource=\regexp{$MAPLOOP}, match={Evolutionary}, replace={Evol.}]
			\step[fieldsource=\regexp{$MAPLOOP}, match={Education}, replace={Educ.}]
			\step[fieldsource=\regexp{$MAPLOOP}, match={Exhibition}, replace={Exhib.}]
			\step[fieldsource=\regexp{$MAPLOOP}, match={Electrical}, replace={Elect.}]
			\step[fieldsource=\regexp{$MAPLOOP}, match={Electric}, replace={Elect.}]
			\step[fieldsource=\regexp{$MAPLOOP}, match={Experimental}, replace={Exp.}]
			\step[fieldsource=\regexp{$MAPLOOP}, match={Electrification}, replace={Electrific.}]
			\step[fieldsource=\regexp{$MAPLOOP}, match={Exploratory}, replace={Explor.}]
			\step[fieldsource=\regexp{$MAPLOOP}, match={Electromagnetic}, replace={Electromagn.}]
			\step[fieldsource=\regexp{$MAPLOOP}, match={Exposition}, replace={Expo.}]
			\step[fieldsource=\regexp{$MAPLOOP}, match={Electroacoustic}, replace={Electroacoust.}]
			\step[fieldsource=\regexp{$MAPLOOP}, match={Express}, replace={Express}]
			\step[fieldsource=\regexp{$MAPLOOP}, match={Electronics}, replace={Electron.}]
			\step[fieldsource=\regexp{$MAPLOOP}, match={Electronic}, replace={Electron.}]
			\step[fieldsource=\regexp{$MAPLOOP}, match={Fabrication}, replace={Fabr.}]
			\step[fieldsource=\regexp{$MAPLOOP}, match={Emerging}, replace={Emerg.}]
			\step[fieldsource=\regexp{$MAPLOOP}, match={Faculty}, replace={Fac.}]
			\step[fieldsource=\regexp{$MAPLOOP}, match={Engineering}, replace={Eng.}]
			\step[fieldsource=\regexp{$MAPLOOP}, match={Engineers}, replace={Eng.}]
			\step[fieldsource=\regexp{$MAPLOOP}, match={Engineer}, replace={Eng.}]
			\step[fieldsource=\regexp{$MAPLOOP}, match={Ferroelectrics}, replace={Ferroelect.}]
			\step[fieldsource=\regexp{$MAPLOOP}, match={Environment}, replace={Environ.}]
			\step[fieldsource=\regexp{$MAPLOOP}, match={Francais, French}, replace={Fr.}]
			\step[fieldsource=\regexp{$MAPLOOP}, match={Equations}, replace={Equ.}]
			\step[fieldsource=\regexp{$MAPLOOP}, match={Frequency}, replace={Freq.}]
			\step[fieldsource=\regexp{$MAPLOOP}, match={Equipment}, replace={Equip.}]
			\step[fieldsource=\regexp{$MAPLOOP}, match={Foundation}, replace={Found.}]
			\step[fieldsource=\regexp{$MAPLOOP}, match={Ergonomics}, replace={Ergonom.}]
			\step[fieldsource=\regexp{$MAPLOOP}, match={Fundamental}, replace={Fundam.}]
			\step[fieldsource=\regexp{$MAPLOOP}, match={European}, replace={Eur.}]
			\step[fieldsource=\regexp{$MAPLOOP}, match={Generation}, replace={Gener.}]
			\step[fieldsource=\regexp{$MAPLOOP}, match={Evaluation}, replace={Eval.}]
			\step[fieldsource=\regexp{$MAPLOOP}, match={Geology}, replace={Geol.}]
			\step[fieldsource=\regexp{$MAPLOOP}, match={Geophysics}, replace={Geophys.}]
			\step[fieldsource=\regexp{$MAPLOOP}, match={Innovations}, replace={Innov.}]
			\step[fieldsource=\regexp{$MAPLOOP}, match={Innovation}, replace={Innov.}]
			\step[fieldsource=\regexp{$MAPLOOP}, match={Geoscience}, replace={Geosci.}]
			\step[fieldsource=\regexp{$MAPLOOP}, match={Institute}, replace={Inst.}]
			\step[fieldsource=\regexp{$MAPLOOP}, match={Institution}, replace={Inst.}]
			\step[fieldsource=\regexp{$MAPLOOP}, match={Graphics}, replace={Graph.}]
			\step[fieldsource=\regexp{$MAPLOOP}, match={Instrument}, replace={Instrum.}]
			\step[fieldsource=\regexp{$MAPLOOP}, match={Guidance}, replace={Guid.}]
			\step[fieldsource=\regexp{$MAPLOOP}, match={Instrumentation}, replace={Instrum.}]
			\step[fieldsource=\regexp{$MAPLOOP}, match={Harmonics}, replace={Harmon.}]
			\step[fieldsource=\regexp{$MAPLOOP}, match={Harmonic}, replace={Harmon.}]
			\step[fieldsource=\regexp{$MAPLOOP}, match={Insulation}, replace={Insul.}]
			\step[fieldsource=\regexp{$MAPLOOP}, match={History}, replace={Hist.}]
			\step[fieldsource=\regexp{$MAPLOOP}, match={Integrated}, replace={Integr.}]
			\step[fieldsource=\regexp{$MAPLOOP}, match={Horizon}, replace={Horiz.}]
			\step[fieldsource=\regexp{$MAPLOOP}, match={Intelligence}, replace={Intell.}]
			\step[fieldsource=\regexp{$MAPLOOP}, match={Hungary}, replace={Hung.}]
			\step[fieldsource=\regexp{$MAPLOOP}, match={Hungarian}, replace={Hung.}]
			\step[fieldsource=\regexp{$MAPLOOP}, match={Intelligent}, replace={Intell.}]
			\step[fieldsource=\regexp{$MAPLOOP}, match={Hydraulics}, replace={Hydraul.}]
			\step[fieldsource=\regexp{$MAPLOOP}, match={Interactions}, replace={Interact.}]
			\step[fieldsource=\regexp{$MAPLOOP}, match={Hydrology}, replace={Hydrol.}]
			\step[fieldsource=\regexp{$MAPLOOP}, match={Internationales}, replace={Int.}]
			\step[fieldsource=\regexp{$MAPLOOP}, match={International}, replace={Int.}]
			\step[fieldsource=\regexp{$MAPLOOP}, match={Illuminating}, replace={Illum.}]
			\step[fieldsource=\regexp{$MAPLOOP}, match={Isotopes}, replace={Isot.}]
			\step[fieldsource=\regexp{$MAPLOOP}, match={Imaging}, replace={Imag.}]
			\step[fieldsource=\regexp{$MAPLOOP}, match={Israel}, replace={Isr.}]
			\step[fieldsource=\regexp{$MAPLOOP}, match={Industrial}, replace={Ind.}]
			\step[fieldsource=\regexp{$MAPLOOP}, match={Japan}, replace={Jpn.}]
			\step[fieldsource=\regexp{$MAPLOOP}, match={Information}, replace={Inf.}]
			\step[fieldsource=\regexp{$MAPLOOP}, match={Journal}, replace={J.}]
			\step[fieldsource=\regexp{$MAPLOOP}, match={Informatics}, replace={Inform.}]
			\step[fieldsource=\regexp{$MAPLOOP}, match={Knowledge}, replace={Knowl.}]
			\step[fieldsource=\regexp{$MAPLOOP}, match={Laboratory}, replace={Lab.}]
			\step[fieldsource=\regexp{$MAPLOOP}, match={Laboratories}, replace={Lab.}]
			\step[fieldsource=\regexp{$MAPLOOP}, match={Mathematical}, replace={Math.}]
			\step[fieldsource=\regexp{$MAPLOOP}, match={Language}, replace={Lang.}]
			\step[fieldsource=\regexp{$MAPLOOP}, match={Mathematics}, replace={Math.}]
			\step[fieldsource=\regexp{$MAPLOOP}, match={Abstracts}, replace={Abstr.}]
			\step[fieldsource=\regexp{$MAPLOOP}, match={Analysis}, replace={Anal.}]
			\step[fieldsource=\regexp{$MAPLOOP}, match={Academy}, replace={Acad.}]
			\step[fieldsource=\regexp{$MAPLOOP}, match={Annals}, replace={Ann.}]
			\step[fieldsource=\regexp{$MAPLOOP}, match={Accelerator}, replace={Accel.}]
			\step[fieldsource=\regexp{$MAPLOOP}, match={Annual}, replace={Annu.}]
			\step[fieldsource=\regexp{$MAPLOOP}, match={Acoustics}, replace={Acoust.}]
			\step[fieldsource=\regexp{$MAPLOOP}, match={Apparatus}, replace={App.}]
			\step[fieldsource=\regexp{$MAPLOOP}, match={Active}, replace={Act.}]
			\step[fieldsource=\regexp{$MAPLOOP}, match={Applications}, replace={Appl.}]
			\step[fieldsource=\regexp{$MAPLOOP}, match={Administration}, replace={Admin.}]
			\step[fieldsource=\regexp{$MAPLOOP}, match={Applied}, replace={Appl.}]
			\step[fieldsource=\regexp{$MAPLOOP}, match={Administrative}, replace={Administ.}]
			\step[fieldsource=\regexp{$MAPLOOP}, match={Approximate}, replace={Approx.}]
			\step[fieldsource=\regexp{$MAPLOOP}, match={Advanced}, replace={Adv.}]
			\step[fieldsource=\regexp{$MAPLOOP}, match={Advances}, replace={Adv.}]
			\step[fieldsource=\regexp{$MAPLOOP}, match={Archives}, replace={Arch.}]
			\step[fieldsource=\regexp{$MAPLOOP}, match={Archive}, replace={Arch.}]
			\step[fieldsource=\regexp{$MAPLOOP}, match={Aeronautics}, replace={Aeronaut.}]
			\step[fieldsource=\regexp{$MAPLOOP}, match={Artificial}, replace={Artif.}]
			\step[fieldsource=\regexp{$MAPLOOP}, match={Aerospace}, replace={Aerosp.}]
			\step[fieldsource=\regexp{$MAPLOOP}, match={Assembly}, replace={Assem.}]
			\step[fieldsource=\regexp{$MAPLOOP}, match={Affective}, replace={Affect.}]
			\step[fieldsource=\regexp{$MAPLOOP}, match={Association}, replace={Assoc.}]
			\step[fieldsource=\regexp{$MAPLOOP}, match={Africa}, replace={Afr.}]
			\step[fieldsource=\regexp{$MAPLOOP}, match={African}, replace={Afr.}]
			\step[fieldsource=\regexp{$MAPLOOP}, match={Astronomy}, replace={Astron.}]
			\step[fieldsource=\regexp{$MAPLOOP}, match={Aircraft}, replace={Aircr.}]
			\step[fieldsource=\regexp{$MAPLOOP}, match={Astronautics}, replace={Astronaut.}]
			\step[fieldsource=\regexp{$MAPLOOP}, match={Algebraic}, replace={Algebr.}]
			\step[fieldsource=\regexp{$MAPLOOP}, match={Astrophysics}, replace={Astrophys.}]
			\step[fieldsource=\regexp{$MAPLOOP}, match={American}, replace={Amer.}]
			\step[fieldsource=\regexp{$MAPLOOP}, match={Atmosphere}, replace={Atmos.}]
			\step[fieldsource=\regexp{$MAPLOOP}, match={Atomic}, replace={At.}]
			\step[fieldsource=\regexp{$MAPLOOP}, match={Atoms}, replace={At.}]
			\step[fieldsource=\regexp{$MAPLOOP}, match={Broadcasting}, replace={Broadcast.}]
			\step[fieldsource=\regexp{$MAPLOOP}, match={Australasian}, replace={Australas.}]
			\step[fieldsource=\regexp{$MAPLOOP}, match={Bulletin}, replace={Bull.}]
			\step[fieldsource=\regexp{$MAPLOOP}, match={Australia}, replace={Aust.}]
			\step[fieldsource=\regexp{$MAPLOOP}, match={Bureau}, replace={Bur.}]
			\step[fieldsource=\regexp{$MAPLOOP}, match={{Automatic }}, replace={Autom.}]
			\step[fieldsource=\regexp{$MAPLOOP}, match={Business}, replace={Bus.}]
			\step[fieldsource=\regexp{$MAPLOOP}, match={Automation}, replace={Automat.}]
			\step[fieldsource=\regexp{$MAPLOOP}, match={Canadian}, replace={Can.}]
			\step[fieldsource=\regexp{$MAPLOOP}, match={Automotive}, replace={Automot.}]
			\step[fieldsource=\regexp{$MAPLOOP}, match={Automobiles}, replace={Automob.}]
			\step[fieldsource=\regexp{$MAPLOOP}, match={Automobile}, replace={Automob.}]
			\step[fieldsource=\regexp{$MAPLOOP}, match={Ceramic}, replace={Ceram.}]
			\step[fieldsource=\regexp{$MAPLOOP}, match={Autonomous}, replace={Auton.}]
			\step[fieldsource=\regexp{$MAPLOOP}, match={Chemical}, replace={Chem.}]
			\step[fieldsource=\regexp{$MAPLOOP}, match={Behavioral}, replace={Behav.}]
			\step[fieldsource=\regexp{$MAPLOOP}, match={Behavior}, replace={Behav.}]
			\step[fieldsource=\regexp{$MAPLOOP}, match={Chinese}, replace={Chin.}]
			\step[fieldsource=\regexp{$MAPLOOP}, match={Belgian}, replace={Belg.}]
			\step[fieldsource=\regexp{$MAPLOOP}, match={Climatology}, replace={Climatol.}]
			\step[fieldsource=\regexp{$MAPLOOP}, match={Biochemical}, replace={Biochem.}]
			\step[fieldsource=\regexp{$MAPLOOP}, match={Clinical}, replace={Clin.}]
			\step[fieldsource=\regexp{$MAPLOOP}, match={Bioinformatics}, replace={Bioinf.}]
			\step[fieldsource=\regexp{$MAPLOOP}, match={Cognitive}, replace={Cogn.}]
			\step[fieldsource=\regexp{$MAPLOOP}, match={Biology, Biological}, replace={Biol.}]
			\step[fieldsource=\regexp{$MAPLOOP}, match={Colloquium}, replace={Colloq.}]
			\step[fieldsource=\regexp{$MAPLOOP}, match={Kolloquium}, replace={Kolloq.}]
			\step[fieldsource=\regexp{$MAPLOOP}, match={Biomedical}, replace={Biomed.}]
			\step[fieldsource=\regexp{$MAPLOOP}, match={Communications}, replace={Commun.}]
			\step[fieldsource=\regexp{$MAPLOOP}, match={Communication}, replace={Commun.}]
			\step[fieldsource=\regexp{$MAPLOOP}, match={Biophysics}, replace={Biophys.}]
			\step[fieldsource=\regexp{$MAPLOOP}, match={Compatibility}, replace={Compat.}]
			\step[fieldsource=\regexp{$MAPLOOP}, match={British}, replace={Brit.}]
			\step[fieldsource=\regexp{$MAPLOOP}, match={Components}, replace={Compon.}]
			\step[fieldsource=\regexp{$MAPLOOP}, match={Component}, replace={Compon.}]
			\step[fieldsource=\regexp{$MAPLOOP}, match={Computational}, replace={Comput.}]
			\step[fieldsource=\regexp{$MAPLOOP}, match={Delivery}, replace={Del.}]
			\step[fieldsource=\regexp{$MAPLOOP}, match={Computers}, replace={Comput.}]
			\step[fieldsource=\regexp{$MAPLOOP}, match={Computer}, replace={Comput.}]
			\step[fieldsource=\regexp{$MAPLOOP}, match={Department}, replace={Dept.}]
			\step[fieldsource=\regexp{$MAPLOOP}, match={Computing}, replace={Comput.}]
			\step[fieldsource=\regexp{$MAPLOOP}, match={Design}, replace={Des.}]
			\step[fieldsource=\regexp{$MAPLOOP}, match={Condensed}, replace={Condens.}]
			\step[fieldsource=\regexp{$MAPLOOP}, match={Detector}, replace={Detect.}]
			\step[fieldsource=\regexp{$MAPLOOP}, match={Conferences}, replace={Conf.}]
			\step[fieldsource=\regexp{$MAPLOOP}, match={Conference}, replace={Conf.}]
			\step[fieldsource=\regexp{$MAPLOOP}, match={Development}, replace={Develop.}]
			\step[fieldsource=\regexp{$MAPLOOP}, match={Congress}, replace={Congr.}]
			\step[fieldsource=\regexp{$MAPLOOP}, match={Differential}, replace={Differ.}]
			\step[fieldsource=\regexp{$MAPLOOP}, match={Consumer}, replace={Consum.}]
			\step[fieldsource=\regexp{$MAPLOOP}, match={Digest}, replace={Dig.}] 
			\step[fieldsource=\regexp{$MAPLOOP}, match={{ of the }}, replace={{ }}]
			\step[fieldsource=\regexp{$MAPLOOP}, match={{ of }}, replace={{ }}]
			\step[fieldsource=\regexp{$MAPLOOP}, match={{Of }}, replace={{ }}]
			\step[fieldsource=\regexp{$MAPLOOP}, match={{ on }}, replace={{ }}]
			\step[fieldsource=\regexp{$MAPLOOP}, match={{ On }}, replace={{ }}]
			\step[fieldsource=\regexp{$MAPLOOP}, match={{On }}, replace={{ }}]
			\step[fieldsource=\regexp{$MAPLOOP}, match={{ in }}, replace={{ }}]
			\step[fieldsource=\regexp{$MAPLOOP}, match=\regexp{\\\x{26}}, replace={{and}}] 
			\step[fieldsource=\regexp{$MAPLOOP}, match={{, and }}, replace={{, }}]
			\step[fieldsource=\regexp{$MAPLOOP}, match={{, And }}, replace={{, }}]
			\step[fieldsource=\regexp{$MAPLOOP}, match={{ and }}, replace={{ }}]
			\step[fieldsource=\regexp{$MAPLOOP}, match={{ And }}, replace={{ }}]
			\step[fieldsource=\regexp{$MAPLOOP}, match={{ In }}, replace={{ }}]
			\step[fieldsource=\regexp{$MAPLOOP}, match={{In }}, replace={{ }}]
			\step[fieldsource=\regexp{$MAPLOOP}, match={{ the }}, replace={{ }}] 
			\step[fieldsource=\regexp{$MAPLOOP}, match={{ The }}, replace={{ }}] 
			\step[fieldsource=\regexp{$MAPLOOP}, match={{The }}, replace={}] 
			\step[fieldsource=\regexp{$MAPLOOP}, match={{ for }}, replace={{ }}] 
			\step[fieldsource=\regexp{$MAPLOOP}, match={First}, replace={1st}]
			\step[fieldsource=\regexp{$MAPLOOP}, match={Second}, replace={2nd}]
			\step[fieldsource=\regexp{$MAPLOOP}, match={Third}, replace={3rd}]
			\step[fieldsource=\regexp{$MAPLOOP}, match={Fourth}, replace={4th}]
			\step[fieldsource=\regexp{$MAPLOOP}, match={Fifth}, replace={5th}]
			\step[fieldsource=\regexp{$MAPLOOP}, match={Sixth}, replace={6th}]
			\step[fieldsource=\regexp{$MAPLOOP}, match={Seventh}, replace={7th}]
			\step[fieldsource=\regexp{$MAPLOOP}, match={Eighth}, replace={8th}]
			\step[fieldsource=\regexp{$MAPLOOP}, match={Ninth}, replace={9th}]
			\step[fieldsource=\regexp{$MAPLOOP}, match={Tenth}, replace={10th}]
			\step[fieldsource=\regexp{$MAPLOOP}, match={Eleventh}, replace={11th}]
			\step[fieldsource=\regexp{$MAPLOOP}, match={Twelfth}, replace={12th}]
			\step[fieldsource=\regexp{$MAPLOOP}, match={Thirteenth}, replace={13th}]
			\step[fieldsource=\regexp{$MAPLOOP}, match={Fourteenth}, replace={14th}]
			\step[fieldsource=\regexp{$MAPLOOP}, match={Fifteenth}, replace={15th}]
			\step[fieldsource=\regexp{$MAPLOOP}, match={Sixteenth}, replace={16th}]
			\step[fieldsource=\regexp{$MAPLOOP}, match={Seventeenth}, replace={17th}]
			\step[fieldsource=\regexp{$MAPLOOP}, match={Eighteenth}, replace={18th}]
			\step[fieldsource=\regexp{$MAPLOOP}, match={Nineteenth}, replace={19th}]
			\step[fieldsource=\regexp{$MAPLOOP}, match={Twentieth}, replace={20th}]
		}
	}
}

\usepackage{xpatch}
\usepackage{xstring}

\DeclareSourcemap{
	\maps[datatype=bibtex]{
		\map{
			\step[fieldsource=langid, match=\regexp{\A(n)?((swiss)?german|austrian)\Z}, final]
			\step[fieldset=language, fieldvalue={(in German)}]
		}
		\map{
			\step[fieldsource=langid, match=pinyin, final]
			\step[fieldset=language, fieldvalue={(in Chinese)}]
		}
		\map{
			\step[fieldsource=langid, match=japanese, final]
			\step[fieldset=language, fieldvalue={(in Japanese)}]
		}
		\map{
			\step[fieldsource=langid, match=ko, final]
			\step[fieldset=language, fieldvalue={(in Korean)}]
		}
	}
}

\newbibmacro*{publisher+location}{%
	\printlist{location}%
	\iflistundef{publisher}
	{\setunit*{\addcomma\space}}
	{\setunit*{\addcolon\space}}%
	\printlist{publisher}%
	\newunit}

\xpatchbibdriver{online}
{\printtext[parens]{\usebibmacro{date}}}
	{\iffieldundef{year}
		{}
		{\printtext[parens]{\usebibmacro{date}}}}
	{}
	{\typeout{There was an error patching biblatex-ieee (specifically, ieee.bbx's @online driver)}}

\DeclareSourcemap{
	\maps[datatype=biber]{
		\map{
			\step[fieldsource=note, final]
			\step[fieldset=addendum, origfieldval, final]
			\step[fieldset=note, null]
		}
	}
}

\DeclareBibliographyDriver{standard}{%
	\usebibmacro{bibindex}%
	\usebibmacro{begentry}%
	\usebibmacro{maintitle+title}%
	\setunit{\addcomma\addspace}%
	\usebibmacro{publisher+type+number}%
	\setunit{\labelnamepunct}\newblock
	\IfSubStr{\strfield{number}}{\strfield{year}}{}{\usebibmacro{date}} 
	\newunit\newblock
	\usebibmacro{finentry}%
}

\newbibmacro*{publisher+type+number}{%
	\printtext{%
		\printlist{publisher}
		\iflistundef{publisher}
		{\space}{}%
		\iffieldundef{type}
		{Standard}
		{\printfield{type}}
		\printfield{number}
	}%
}

\xpatchbibdriver{report}					 	
	{\usebibmacro{date}}			
	{\IfSubStr{\strfield{number}}{\strfield{year}}{}{\usebibmacro{date}}} 
	{} 
	{\typeout{There was an error patching biblatex-ieee (specifically, ieee.bbx's @report driver)}} 

\DeclareBibliographyDriver{software}{%
	\usebibmacro{bibindex}%
	\usebibmacro{begentry}%
	\usebibmacro{maintitle+title}%
	\setunit{\addcomma\addspace}%
	\usebibmacro{version+year}%
	\setunit{\labelnamepunct}\newblock
	\usebibmacro{author}%
	\setunit{\addcomma\addspace}%
	\newunit\newblock
	\usebibmacro{url+urldate}%
	\newunit\newblock
	\usebibmacro{finentry}%
}

\newbibmacro*{version+year}{%
	\printtext{%
		\iffieldundef{version}{\iffieldundef{year}{}{\printfield{year}}}{\printfield{version}}
		\printfield{number}
		\printlist{publisher}
		\iflistundef{publisher}{\space}{}%
	}%
}

\usepackage[T1]{fontenc}

\usepackage{hyperref}

\usepackage{graphicx}

\usepackage{amsmath}

\usepackage{cleveref}

\usepackage{tabularx}

\usepackage{ifthen}

\usepackage{tikz}
\usetikzlibrary{arrows.meta}

\newcommand{\withcoverpage}{1}
\newcommand{\authorstring}{Niklas Braun, Markus Steimle, Martin Törngren, and Markus Maurer}
\newcommand{\authorstringaff}{Niklas Braun$^{1}$, Markus Steimle$^{1}$, Martin Törngren$^{2}$, and Markus Maurer$^{1}$}
\newcommand{\yearstring}{2024}

\newcommand{\titlestring}{A Concept for Semi-Automatic Configuration of Sufficiently Valid Simulation Setups for Automated Driving Systems}

\title{\LARGE \bf
  \titlestring
}

\author{\authorstringaff
  \thanks{This research has been carried out as part of the project \mbox{``AUTOtech.\emph{agil}''} (FKZ 01IS22088R) supported by the German Federal Ministry of Education and Research of Germany (BMBF) and by the TADDO2 project supported by the Swedish innovation agency Vinnova (proj. 2024-00816).}
  \thanks{$^{1}$Niklas Braun, Markus Steimle and Markus Maurer are with the Institute of Control Engineering,
    Technische Universität Braunschweig, 38106 Braunschweig, Germany
      {\tt \{niklas.braun, m.steimle, markus.maurer\}@tu-braunschweig.de}}%
  \thanks{$^{2}$Martin Törngren is with the Department of Engineering Design, KTH Royal Institute of Technology, 10044 Stockholm, Sweden
      {\tt martint@kth.se}}%
}

\ifthenelse{\withcoverpage=1}{
	\usepackage{hologo}
\usepackage{listings}
\lstset{breaklines,basicstyle=\small,columns=fullflexible,basicstyle=\ttfamily,language={[plain]TeX}}
}{}

\begin{document}

\ifthenelse{\withcoverpage=1}{
	\newcommand{\conferencestring}{27th IEEE International Conference on Intelligent Transportation Systems}
\newcommand{\addressstring}{Edmonton, Canada}

\twocolumn[
\begin{@twocolumnfalse}
  \Huge {IEEE copyright notice} \\ \\
	\large {\copyright\ \yearstring\ IEEE. Personal use of this material is permitted. Permission from IEEE must be obtained for all other uses, in any current or future media, including reprinting/republishing this material for advertising or promotional purposes, creating new collective works, for resale or redistribution to servers or lists, or reuse of any copyrighted component of this work in other works.} \\ \\
	
	{\Large Published in \emph{\conferencestring}, \addressstring, September 24--27, 2024.} \\ \\ 

  {\Large DOI: \href{https://doi.org/10.1109/ITSC58415.2024.10919986}{10.1109/ITSC58415.2024.10919986}} \\ \\

Cite as:
\vspace{0.1cm}

\noindent\fbox{%
    \parbox{\textwidth}{%
        \authorstring, ``\titlestring,''
  in \emph{\conferencestring}, \addressstring, \yearstring, pp. 2042--2049, doi: 10.1109/ITSC58415.2024.10919986
    }%
}
\vspace{2cm}

\end{@twocolumnfalse}
]

\noindent\begin{minipage}{\textwidth}

\hologo{BibTeX}:
\footnotesize
\begin{lstlisting}[frame=single]
@inproceedings{braun_simulation_2024,
  author={{Braun}, Niklas and {Steimle}, Markus and {T{\"o}rngren}, Martin and {Maurer}, Markus},
  booktitle={2024 {IEEE} 27th {International} {Conference} on {Intelligent} {Transportation} {Systems} {({ITSC})}},
  title={A {Concept} for {Semi-Automatic} {Configuration} of {Sufficiently} {Valid} {Simulation} {Setups} for {Automated} {Driving} {Systems}},
  address={Edmonton, Canada},
  year={2024},
  pages={2042--2049},
  doi={10.1109/ITSC58415.2024.10919986},
  publisher={IEEE}
}
\end{lstlisting}
\end{minipage}
}{}

\maketitle
\thispagestyle{empty}
\pagestyle{empty}

\begin{abstract}

    As simulation is increasingly used in scenario-based approaches to test Automated Driving Systems, the credibility of simulation results is a major concern.
    Arguably, credibility depends on the validity of the simulation setup and simulation models.
    When selecting appropriate simulation models, a trade-off must be made between validity, often connected to the model's fidelity, and cost of computation.
    However, due to the large number of test cases, expert-based methods to create sufficiently valid simulation setups seem infeasible.
    We propose using design contracts in order to semi-automatically compose simulation setups for given test cases from simulation models and to derive requirements for the simulation models, supporting separation of concerns between simulation model developers and users.
    Simulation model contracts represent their validity domains by capturing a validity guarantee and the associated operating conditions in an assumption.
    We then require the composition of the simulation model contracts to refine a test case contract.
    The latter contract captures the operating conditions of the test case in its assumption and validity requirements in its guarantee.
    Based on this idea, we present a framework that supports the compositional configuration of simulation setups based on the contracts and a method to derive runtime monitors for these simulation setups.

\end{abstract}

\section{INTRODUCTION}

An important challenge for the widespread market introduction of vehicles equipped with Automated Driving Systems is estimating the residual risk such systems introduce.
Testing can provide evidence about how a system behaves under certain conditions.
When it comes to testing Automated Driving Systems in particular, simulation is becoming increasingly ubiquitous.
However, the credibility of simulation results is an important concern from the perspective of certain stakeholders.
Credibility is ``the quality to elicit belief or trust in [modeling and simulation] results'' \cite{NASASTD7009A};
hence, credibility cannot be objectively produced.
Nevertheless, methods to reason about the simulation's validity can support credibility.

The need for scenario-based and simulation-based testing arises from the complexity of Automated Driving Systems and their operating context.
Earlier automotive verification and validation approaches rely on driving a certain distance under real-world conditions.
However, neither are these approaches adequately scalable in the context of Automated Driving Systems, nor do they cover the relevant scenarios efficiently and systematically (``approval trap'', \cite{wachenfeldReleaseAutonomousVehicles2016}).
In contrast, in scenario-based approaches, the operational context of the system is structured by defined scenarios for considerations at various stages of the development process \cite{menzelScenariosDevelopmentTestValidation2018}.
Depending on the purpose, scenarios are described on different levels of abstraction, where concrete scenarios are the least abstract.
A scenario consists of actions, events, goals, values, and at least two scenes.
A scene consists of scenery and movable objects \cite{steimleConsistentTaxonomyScenarioBasedDevelopment2021}.
A concrete scenario describes these entities on a parameterized level.
Several approaches to systematic scenario generation and to identifying critical scenarios have been proposed in publications, such as \cite{menzelFunctionalLogicalScenariosDetailing2019,zhangFindingCriticalScenariosAutomated2023}.

The international standard on \textcite{iso15288:2015} defines that validation is the ``confirmation, through the provision of objective evidence, that the requirements for a specific intended use or application have been fulfilled.''
We define testing as a means to provide this evidence.
According to \textcite{steimleConsistentTaxonomyScenarioBasedDevelopment2021}, a test case consists of a concrete scenario and at least one evaluation criterion which is or are derived from one or multiple requirements.
Since the number of scenarios and number of requirements may be very large, the number of resulting test cases may be very large, as well.

Several test bench types exist for executing test cases, for example, Hardware-in-the-Loop (HiL), Software-in-the-Loop (SiL), and real-world driving.
Criteria, such as validity, cost, and feasibility, can be used to determine the most suitable type of test bench.
\textcite{steimleGeneratingSufficientlyValidTest2022} proposed a method for systematically assigning test cases to test benches and their specific configuration.
However, this paper will only focus on simulation-based test benches.

Simulation plays a major role in automated driving research and development due to its inherent advantages regarding scalability, repeatability, and the lack of exposing agents involved in the scenario to physical harm.
One example is the reliance on simulation throughout the development process of a connected and automated mobility system within the research project \mbox{AUTOtech.\emph{agil}} \cite{vankempenAUTOtechAgilArchitectureTechnologies2023}.
To address the validity-related credibility concerns, firstly, the validity of the used simulation models needs to be considered.
Secondly, coupling effects between simulation models and effects of the simulation software may influence the validity of the simulation results.

With this work, we aim to support creating a ``sufficiently valid'' \cite{balciGoldenRulesVerificationValidation2010} test bench configuration for each test case.
We define the validity domain of a simulation model as the set of operating conditions under which the simulation model is sufficiently valid.
As the simulation model is used in a certain context, for example, during a test case execution, the specified validity can be expected to be satisfied if the operating conditions are ensured to be within the validity domain.
By formalizing validity domains using design contracts, we aim to support reasoning about the validity of the simulation setup and hence the test case execution.
We acknowledge the additional challenges introduced by this approach, such as the cost of high dimensionality in formal considerations. Still, we propose this framework as a means to potentially manage complexity in the field of simulation for Automated Driving Systems and to stimulate a discussion about using design contracts in this context.

The remainder of this paper is structured as follows.
In \cref{sec:related-work}, related work regarding scenario-based testing, simulation-based testing, and the conceptualization of simulation models is presented.
\Cref{sec:contract-theory} introduces the contract theory to be applied in this paper based on literature related to assume-guarantee contracts.
In \cref{sec:contracts-for-simulation}, we explain how contracts can specify simulation models' validity domains and test cases.
\Cref{sec:framework} describes a framework that applies these contracts to support the configuration of simulation setups and generation of monitors, and \cref{sec:conclusion} concludes the paper and provides suggestions for future work.
\section{RELATED WORK}
\label{sec:related-work}

\subsection{Scenario-Based Testing}

Scenarios for automated driving are a widely adopted concept to structure an Automated Driving System's operational context.
According to \textcite{ulbrichDefiningSubstantiatingTermsScene2015}, scenarios consist of the temporal development of at least two scenes.
Furthermore, a scenario is characterized by actions and events that occur during the scenario as well as goals and values that are associated with actors in the scenario.
A scene captures the state of the scenery, movable objects, and the actors' and observers' self-representations at a certain point in time.
The scenery can be described in a structured manner, e.g., by using the four-layer model \cite{schuldtEffizienteSystematischeTestgenerierungFur2013}, which was later extended by, for example, \textcite{bagschikOntologyBasedSceneCreation2018,scholtes6LayerModelStructuredDescription2021,wangSurveyEmergingSafetyChallenge2024}.

\textcite{menzelScenariosDevelopmentTestValidation2018} proposed three abstraction levels of scenarios.
Functional scenarios are based on use cases and are described using natural language on a semantic level and with domain-specific vocabulary.
Logical scenarios introduce state variables to the entities of the functional scenarios.
These variables can be described by value ranges or probability distributions.
Finally, concrete scenarios are samples of logical scenarios, specifying the exact values of the state variables in the scenario.
This approach allows tracing concrete scenarios back to use cases and is further detailed in \cite{menzelFunctionalLogicalScenariosDetailing2019,birkemeyerSamplingStrategiesGeneratingScenarios2021}.
A literature review on methods for identifying critical scenarios was presented by \textcite{zhangFindingCriticalScenariosAutomated2023}.

\subsection{Simulation-Based Testing}

For the purposes of this paper, testing is the process of creating objective evidence needed for verification and validation of the system under test.
Based on the taxonomy defined by \textcite{steimleConsistentTaxonomyScenarioBasedDevelopment2021}, a test case for an Automated Driving System or one of its subsystems entails a concrete scenario and at least one evaluation criterion.
The evaluation criterion is derived from a requirement.
Test cases can be executed using various test bench types, e.g., a real vehicle on public roads, a Hardware-in-the-Loop (HiL) test bench, or a Software-in-the-Loop (SiL) test bench.
In the context of simulation, a test bench configuration contains the composition of simulation models along with their parameterizations.

Assigning test cases to test bench configurations is a matter addressed by \textcite{steimleGeneratingSufficientlyValidTest2022}.
\citeauthor{steimleGeneratingSufficientlyValidTest2022} describe a major challenge being to ensure that the test case is executed in a sufficiently valid manner while keeping the effort for test case execution reasonably low.
In their publication, a systematic method to assign test cases to test bench configurations is proposed.
This entails defining characteristics of test bench elements (i.e., simulation models in the context of simulation), such as their validity domain, computational complexity or cost, and interfaces.
In their conclusion, \citeauthor{steimleGeneratingSufficientlyValidTest2022} elaborate on research questions regarding the determining and evaluating simulation models' validity domains and their utilization during the test case assignment process.
We address these questions by the concept proposed in this paper.

\subsection{Simulation Models as Systems}
\label{subsec:simulation-models-as-systems}

Based on \textcite{banksHandbookSimulationPrinciplesMethodology1998}, the \textcite{NASASTD7009A} defines a model as a ``description or representation of a system, entity, phenomenon, or process.''
We distinguish between the conceptual model (as defined in \cite{NASASTD7009A}) and its executable implementation, the simulation model.
An implementation is executable if it can be executed on computer hardware or interpreted by a software, as is common with scripting languages and graphical modeling tools.
A simulation model can exist in various forms, for example, as source code or as compiled binary code.
With this definition, we consider both white-box models where the inner structure is known and black-box models where only the input-output behavior is known.
In the latter case, the simulation model's inner structure is hidden, e.g., for intellectual property reasons.

We describe simulation models as systems with input ports, output ports, and internal states \cite{braunSupportingScenarioBasedValidationAutomated2022thesis}.
A simulation model can be composed of sub-models, resulting in a hierarchy of simulation models.
Simulation models interact with each other and their environment through variables that are associated with their ports.
An output port of a simulation model controls the associated variable's value.
In compliance with the interface theories in \cite{alfaroInterfaceTheoriesComponentBasedDesign2001}, each variable is controlled by exactly one output port;
otherwise, it would be undefined or ambiguously defined.
An input port of a simulation model is uncontrolled by that simulation model.
Parameters of simulation models are represented by input ports that are connected to variables controlled by the simulation platform and that are set to a constant value during the simulation.

\subsection{Simulation Model Validity Domains}

As suggested by \textcite{steimleGeneratingSufficientlyValidTest2022}, the validity domain of a simulation model describes the conditions under which the simulation model satisfies some validity metric to a certain extent.
Due to any model's nature, as described by \textcite{stachowiakAllgemeineModelltheorie1973}, a model cannot be absolutely valid but only ``sufficiently valid'' \cite{balciGoldenRulesVerificationValidation2010} for the application it was validated for.

The concept of ``frames'' for simulation models is described by \textcite{zeiglerTheoryModelingSimulation2000} and was later applied, for instance, by \textcite{vanmierloExploringValidityFramesPractice2020}.
Such frames formalize a simulation model's context, i.e., the operating conditions it is being exposed to during an experiment.
Using that notion, one could define ``validity frames'' for simulation models as the operating conditions under which the simulation model is sufficiently valid.
However, we will use the term ``validity domain'' instead of ``validity frame'' to remain consistent with the terminology used in the main motivating work \cite{steimleGeneratingSufficientlyValidTest2022}.

\subsection{Design Contracts}

Design contracts have been proposed in the field of software engineering, e.g., \cite{meyerApplyingDesignContract1992} and later applied to a broader scope, including the design of cyber-physical systems for supporting component based development, see e.g., \textcite{sangiovanni-vincentelliTamingDrFrankensteinContractBased2012},
as well as to facilitate decoupling and concurrent engineering between CPS disciplines, see e.g., \textcite{derlerCyberphysicalSystemDesign2013}.
For the remainder of this paper, we refer to the term \emph{design contract} as \emph{contract} for brevity.
Several theories about contracts, their applications, and their properties have been elaborated.
Notably, \textcite{benvenisteContractsSystemDesign2018} put forth a meta-theory of contracts based on previous work and a requirements analysis with contracts' applications in cyber-physical system design in mind.
We comply with this meta-theory for the purposes of this paper.
Primarily, we use definitions from \cite{benvenisteMultipleViewpointContractBased2008,westmanConditionsContractsSeparatingResponsibilities2017,westmanStructuringSafetyRequirements2013,incerromeoQuotientAssumeGuaranteeContracts2018,nuzzoPlatformBasedDesignMethodologyContracts2015}.
These publications are referred to in detail in the following section.
\section{Contract Theory}
\label{sec:contract-theory}

We consider the configuration of sufficiently valid simulation setups as a compositional problem where, especially given the scale at which it is applied, the systematic and adequate re-use of components seems desirable.
Aside from the technical challenges, this presents an organizational challenge where responsibilities may be separated, for example, between simulation model developers and simulation model users.
Contract-based design has emerged as an approach to address tasks of this nature.
In this section, we present a contract theory based on literature related to assume-guarantee contracts that we apply later in the context of simulation models and test cases that are to be executed in a simulation environment.

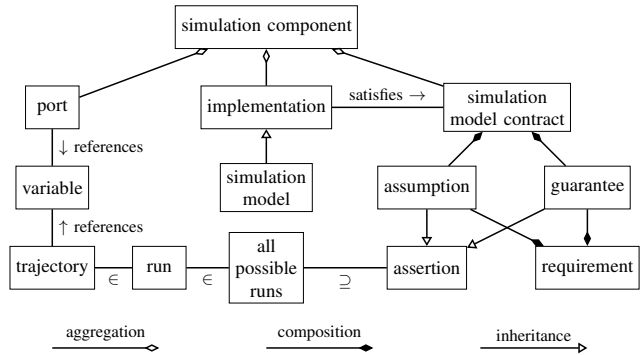
\begin{figure}
    \centering
    \resizebox{\linewidth}{!}{
        \begin{tikzpicture}

  \tikzstyle{taxonomy_rect} = [
    rectangle, 
    draw, 
    fill=white,
    minimum width=1cm, 
    minimum height=0.8cm, 
    align=center
  ];

  \tikzstyle{taxonomy_rect_ref} = [
    rectangle, 
    draw, 
    fill=lightgray, 
    fill opacity=0.7, 
    text opacity=1.0,
    minimum width=1cm, 
    minimum height=0.8cm, 
    align=center
  ];
  
  \tikzstyle{uml_composition} = [
    thick, 
    ->, 
    >=Diamond
    ];
  \tikzstyle{uml_aggregation} = [
    thick, 
    ->, 
    >={Diamond[open]}
    ];
  \tikzstyle{uml_inheritance} = [
    thick, 
    ->, 
    >={Triangle[open]}
    ];
  \tikzstyle{arrow} = [
    thick, 
    ->, 
    >=latex
    ];
  \tikzstyle{no_arrow} = [
    thick
  ];
  
  \node[taxonomy_rect] (component) at (0,3) {simulation component};

  \node[taxonomy_rect] (implementation) at (0,1.5) {implementation};
  \node[taxonomy_rect] (sm) at (0,0) {simulation \\ model};
  \draw[uml_aggregation] (implementation) -- (component);
  \draw[uml_inheritance] (sm) -- (implementation);
  
  \node[taxonomy_rect] (port) at (-4,1.5) {port};
  \draw[uml_aggregation] (port) -- (component);
  
  \node[taxonomy_rect] (contract) at (4.5,1.5) {simulation \\ model contract};
  \node[taxonomy_rect] (assumption) at (3,0) {assumption};
  \node[taxonomy_rect] (guarantee) at (6,0) {guarantee};
  \node[taxonomy_rect] (requirement) at (6,-1.5) {requirement};
  \draw[uml_aggregation] (contract) -- (component);
  \draw[uml_composition] (assumption) -- (contract);
  \draw[uml_composition] (guarantee) -- (contract);
  \draw[uml_composition] (assumption) -- (requirement);
  \draw[uml_composition] (guarantee) -- (requirement);
  \draw[no_arrow] (contract) -- (implementation) node[midway, anchor=south, font=\small] {satisfies $\rightarrow$};

  \node[taxonomy_rect] (assertion) at (3,-1.5) {assertion};
  \draw[uml_inheritance] (assumption) -- (assertion);
  \draw[uml_inheritance] (guarantee) -- (assertion);

  \node[taxonomy_rect] (all_runs) at (0,-1.5) {all \\ possible \\ runs};
  \node[taxonomy_rect] (run) at (-2,-1.5) {run};
  \node[taxonomy_rect] (trajectory) at (-4,-1.5) {trajectory};
  \node[taxonomy_rect] (variable) at (-4,0) {variable};

  \draw[no_arrow] (assertion) -- (all_runs) node[midway, anchor=north, font=\small] {$\supseteq$};
  \draw[no_arrow] (run) -- (all_runs) node[midway, anchor=north, font=\small] {$\in$};
  \draw[no_arrow] (trajectory) -- (run) node[midway, anchor=north, font=\small] {$\in$};
  \draw[no_arrow] (variable) -- (trajectory) node[midway, anchor=west, font=\small, align=center] {$\uparrow$ references};

  \draw[no_arrow] (port) -- (variable) node[midway, anchor=west, font=\small] {$\downarrow$ references};

  \draw[uml_aggregation] (-4,-3) -- (-2,-3) node[midway, anchor=south, font=\small] {aggregation};
  \draw[uml_composition] (0,-3) -- (2,-3) node[midway, anchor=south, font=\small] {composition};
  \draw[uml_inheritance] (4,-3) -- (6,-3) node[midway, anchor=south, font=\small] {inheritance};

  \coordinate (top) at (0,4);
  \coordinate (bottom) at (0,-3.2);
  \path (top);
  \path (bottom);
  
\end{tikzpicture}
    }
    \caption{Ontology of components, contracts, variables, and related concepts in the context of simulation models, based on \cite{braunSupportingScenarioBasedValidationAutomated2022thesis}.}
    \label{fig:ontology_contracts_for_simulation_models}
\end{figure}

The main concepts explained in this section are depicted in the ontology in \Cref{fig:ontology_contracts_for_simulation_models}.
We want to mention that notations, definitions, and terms may vary slightly among the literature on design contracts.
Although the contract-based design community admits taking some liberty concerning using boolean expressions or set notation or even combining the two, we aim to be as consistent in using the set notation as possible.

\subsection{Component}

A component ``represents a unit of design'' \cite{benvenisteMultipleViewpointContractBased2008} that entails an interface, namely the set of input and output ports, and a behavior.
Similarly to \textcite{westmanStructuringSafetyRequirements2013}, we define a component ${I = (\mathcal{P}, \mathcal{M}_{\text{tot}}, \mathcal{C}_{\text{tot}})}$.
\begin{itemize}
    \item ${\mathcal{P} = (P_u, P_c)}$ is the pair of sets of ports of the component $I$. The uncontrolled and controlled ports reference the uncontrolled (input) and controlled (output) variables, respectively.
    \item ${\mathcal{M}_{\text{tot}}}$ is the set of possible implementations of the component $I$. Note that the term \emph{implementation} is potentially ambiguous, as it refers to an entity in ${\mathcal{M}_{\text{tot}}}$ as well as the process of creating that entity. Furthermore, the term refers to a relationship that will be defined in \Cref{subsec:relations-and-properties}. Using different terms wouldn't adequately reflect its use in related work; therefore, we will give context or disambiguation where needed \cite{braunSupportingScenarioBasedValidationAutomated2022thesis}.
    \item ${\mathcal{C}_{\text{tot}}}$ is the set of contracts of the component.
\end{itemize}
\textcite{westmanStructuringSafetyRequirements2013} define a set of sub-components $\mathcal{I}$ as part of the component definition, which we will not use in this paper. We consider the hierarchy implied by this set of sub-components to be flattened, so that the leaf components end up in a parallel composition.

\subsection{Variable and Assertion}

A variable is a quantity that can change over time.
Using the notation from \cite{westmanStructuringSafetyRequirements2013} and \cite{westmanConditionsContractsSeparatingResponsibilities2017}, we define a \emph{trajectory} as the set of pairs ${\{(t, x_i(t))\}_{t \in T}}$, where $T$ is the time domain, and ${x_i(t)}$ is the value of the variable $x_i$ at time $t$.
A set of trajectories of the variables ${x_i \in X}$ is called a \emph{run} of $X$ over $T$.
This definition corresponds to the more colloquial use of the term ``run'' in the context of simulation.
\Textcite{westmanConditionsContractsSeparatingResponsibilities2017} define a set of all runs $\Omega$.

An assertion is a subset ${E \subseteq \Omega}$ and hence constrains the trajectories.
As mentioned previously, there exist boolean notations that express assertions in a semantically equivalent to set notations manner.
The boolean notation is an expression that evaluates to true for a set of trajectories that satisfy the assertion, i.e., are all elements of the assertion set, and that evaluates to false otherwise.

\subsection{Contract}

As part of the meta-theory, \textcite{benvenisteContractsSystemDesign2018} define a contract abstractly as a pair ${\mathcal{C} = (\mathcal{E}_{\mathcal{C}}, \mathcal{M}_{\mathcal{C}})}$.
\begin{itemize}
    \item $\mathcal{E}_{\mathcal{C}}$ is the set of environments that enable the component's operation.
    \item $\mathcal{M}_{\mathcal{C}}$ is the set of implementations that, when operated in an environment in $\mathcal{E}_{\mathcal{C}}$, satisfy the contract.
\end{itemize}
The \citeauthor{benvenisteContractsSystemDesign2018} define that both sets are subsets of a universe $\mathcal{M}$.
Hence, one component's environment can be regarded as a component itself which is a useful consideration in the context of simulation.

Refining the abstract contract definition and using the definitions of trajectories and runs, the assume-guarantee contract is defined as a pair of assertions ${C = (A, G)}$, where $A$ is the assumption and $G$ is the guarantee.
By definition, $A$ constrains the environment and $G$ constrains the component that the contract relates to.
Given a contract, we can derive a requirement toward the component as ${A \implies G}$.
\cite{benvenisteContractsSystemDesign2018}

\subsection{Relations and Properties}
\label{subsec:relations-and-properties}

\subsubsection{Satisfaction and Implementation}

Satisfaction is a relationship between an implementation entity $M$ of a component $I$ and a contract $C$.
$M$ is characterized by its ports $P_M$, and its behavior $B_M$ which is expressed as an assertion over the variables associated with $P_M$ \cite{westmanStructuringSafetyRequirements2013}.
An implementation entity is said to satisfy a contract if and only if ${A \cap B_M \subseteq G}$ \cite{westmanStructuringSafetyRequirements2013}.
It is important that the contract ${C = (A, G)}$ is modeled over the ports $P_M$.

Although the satisfaction and implementation relationships express a similar relationship, they serve different purposes.
Determining satisfaction requires a specification of the behavior $B_M$ and hence would be challenging for black-box implementations.
However, if the specification of an implementation entity is given in the form of the requirement ${G \cup \neg A}$, the implementation relation can be used to determine whether the implementation entity satisfies the contract.

\subsubsection{Refinement}

Refinement is a relationship between two contracts, ${C_1 = (A_1, G_1)}$ and ${C_2 = (A_2, G_2)}$.
The refining contract $C_2$ cannot have stronger assumptions or weaker guarantees than the refined contract $C_1$.
Therefore, ${C_2 \preceq C_1}$ (say: $C_2$ refines $C_1$) if and only if ${A_2 \supseteq A_1 \wedge G_2 \subseteq G_1}$.
Refinement is a transitive relationship, i.e., if ${C_2 \preceq C_1}$ and ${C_3 \preceq C_2}$, then ${C_3 \preceq C_1}$.
This fact can become useful in order to determine the substitutability of components that implement contracts with refinement relationships.

\subsubsection{Compatibility and Consistency}

To define compatibility and consistency as unary properties of contracts, \textcite{benvenisteMultipleViewpointContractBased2008} introduce the concept of receptiveness.
An assertion $E$ made over variables in $P$ is said to be $P'$-receptive if it accepts any trajectories for the subset of variables ${P' \subseteq P}$, i.e., the assertion does not care about variables in $P'$.

For a contract ${C = (A, G)}$ that belongs to a component with (from the perspective of the component) controlled ports $P_c$ and uncontrolled ports $P_u$, the contract is compatible if and only if $A$ is $P_c$-receptive, i.e., the assumption doesn't constrain the controlled ports.
The contract is consistent if and only if $G$ is $P_u$-receptive, i.e., the guarantee doesn't constrain the uncontrolled ports.

\Citeauthor{benvenisteMultipleViewpointContractBased2008} further provide another notion of compatibility and consistency that is used as a relationship between contracts.
Two contracts are said to be compatible (or consistent, respectively) if their parallel composition is compatible (or consistent, respectively).

\subsubsection{Composability}

As a relation between two contracts, composability expresses whether the variable types that these contracts reference match.
\Textcite{benvenisteContractsSystemDesign2018} apply this definition to components as well and declare composability as a purely syntactic property on the components' ports.
In the context of simulation models, the syntax of ports can consist, for example, of data types and physical units.

\subsection{Operations}

\subsubsection{Projection and Inverse Projection}
\label{subsubsec:projection}

Let an assertion $E$ be made over variables $X'$ and let there be a subset of variables ${X \subseteq X'}$.
\Textcite{westmanStructuringSafetyRequirements2013} define the projection $\text{proj}_X(E)$ as the assertion made over the variables $X$ that is obtained by removing all variables not in $X$ from $E$.
This process reduces the dimensionality of the assertion;
hence, the projection is a lossy operation.
In turn, the inverse projection extends the set of variables over which an assertion is made.
The projection and inverse projection are not bijective operations.

\subsubsection{Parallel Composition and Quotient}
\label{subsubsec:parallel-composition-and-quotient}

Parallel composition in the context of contracts is a binary operation that takes two contracts of two distinct component instances and returns a new contract.
This contract fulfills both contracts' guarantees.
The composite contract's assumption is the intersection of the individual contracts' assumptions weakened by the negation of the new guarantee.
Illustratively, the guarantee of one contract helps satisfy the other's assumption and vice versa.
The parallel composition of two contracts ${C_1 = (A_1, G_1)}$ and ${C_2 = (A_2, G_2)}$ is defined as ${C_1 \otimes C_2 = (A_1 \cap A_2 \cup \neg (G_1 \cap G_2), G_1 \cap G_2)}$ \cite{nuzzoPlatformBasedDesignMethodologyContracts2015}.
A precondition for the parallel composition is that the contracts are composable, i.e., that the variable types that the contracts reference match.
This step is called \emph{alphabet equalization} and uses the inverse projection operation \cite{benvenisteContractsSystemDesign2018,nuzzoPlatformBasedDesignMethodologyContracts2015}.
It should be noted that, due to the non-bijective nature of the projection operation, the inverse projection should not be reverted without caution.

The dual operation to parallel composition is the quotient operation.
Let ${C_{\text{top}} = (A_{\text{top}}, G_{\text{top}})}$ and ${C_1 = (A_1, G_1)}$ be saturated contracts.
The quotient ${C_2 = C_{\text{top}} / C_1}$ is defined to yield a contract that parallel-composes with $C_1$ to $C_{\text{top}}$.
A contract $C = (A, G)$ can be saturated by $(A, G) \mapsto (A, G \cup \neg A)$, which is a formulation that gives up on the principle of separation of concerns.
More formally, a saturated contract is not consistent as the guarantee is not $P_u$-receptive due to the inclusion of ${\neg A}$.
However, this way, \textcite{incerromeoQuotientAssumeGuaranteeContracts2018} derive the quotient as ${C_2 = (A_{\text{top}} \cap G_1, A_1 \cap G_{\text{top}} \cup \neg (A_{\text{top}} \cap G_1))}$.

\subsubsection{Conjunction}

The conjunction of two contracts relates assumptions and guarantees with regard to the contracts' refinement preorder, which \textcite{benvenisteContractsSystemDesign2018} call the \emph{shared refinement property}.
For the conjunction ${C = C_1 \wedge C_2}$, this property is ${(C \preceq C_1) \wedge (C \preceq C_2)}$.
Illustratively, any implementation of $C$ satisfies both $C_1$ and $C_2$.
This property is useful for representing different viewpoints on the same component.
The conjunction of contracts is defined as ${C = (A_1 \cup A_2, G_1 \cap G_2)}$.
\section{CONTRACTS FOR SIMULATION}
\label{sec:contracts-for-simulation}

Based on the contract definition and associated operations, we propose how contracts can be applied in the context of simulation models and test cases that are to be executed in a simulation environment.
We provide a summary of the contracts described here at the end of the section in \Cref{fig:summary_contracts_test_case_and_sim_model}.

\subsection{Simulation Model Validity Domains}

We define sufficient validity of a simulation model as the condition that certain validity metrics are satisfied (possibly judged based on expert knowledge) under a certain set of operating conditions.
As the validity metric represents how well the simulation model's behavior matches that of its real counterpart in its expected application, the validity metric takes the model's outputs into account.
In contrast, the operating conditions are defined by the simulation model's inputs.
As mentioned previously, this notion is very similar to the work by \textcite{vanmierloExploringValidityFramesPractice2020} who define a ``validity frame'' for simulation models of, in their example, an electrical circuit component.

We define the validity domain of a simulation model as the set of operating conditions under which the simulation model is sufficiently valid.
This supports a separation of responsibilities among simulation model developers and simulation model users.
As the simulation model is used in a certain context, for example, during a test case execution, the specified validity can be expected to be satisfied if the operating conditions are ensured to be within the validity domain.

\subsection{Coupling Effects and Co-Simulation}

In this paper, we assume that the behavior of simulation models is bounded by the definition of their validity domains.
Hence, we assume that the emerging behavior of a composition of simulation models into a simulation setup is predictable by analyzing their validity domains.
However, this assumption is not always valid as the coupling of systems in general can lead to complex and non-predictable emerging behavior.
Furthermore, connecting multiple simulation platforms or tools, e.g., through some software interface, may introduce additional effects that cannot be described on the level of a simulation model.
For example, a network connection between two concurrent physical computers hosting parts of the simulation may introduce effects, such as latency in the variable propagation between simulation models.
This could be accounted for by adapting the execution model of the simulation platform to track propagation delays.
Another mitigation strategy may be introducing a simulation model for such a delay, which, however, remains an open topic for future work.
Instead, we adopt the approach of \textcite{alfaroInterfaceTheoriesComponentBasedDesign2001} and consider the interconnects between simulation models to be functionally transparent.

\subsection{Simulation Model Contracts}

With the contract theory and conceptualization of simulation models described, we relate the concepts to simulation models.
A simulation component, see \Cref{fig:ontology_sim_arch_and_setup}, consists of any number of ports, any number of implementations, and any number of contracts.
An implementation is an abstract entity that is instantiated by a simulation model.
A simulation model has a contract associated with it which represents its validity domain.
The contract's guarantee and assumption constrain the set of possible runs;
hence, they need to reference variables.
These variables are the same variables as referenced by the ports and are part of a common domain model.

\subsection{Test Case Contracts}

The test case contract is constructed based on the test case description.
The aim is to formalize the operating conditions of the system under test and the validity requirements associated with the evaluation criterion.

A test case entails a concrete scenario which includes operating conditions for the system under test.
We formalize the operating conditions as an assertion over a set of variables of a common domain model, to which, as explained above, the simulation models are referenced to as well.
\textcite{menzelFunctionalLogicalScenariosDetailing2019} presented a keyword-based approach that uses a domain ontology to derive that set of variables from a scenario description.
One domain ontology for automated driving that could be used here is the ASAM OpenXOntology \cite{ASAMOpenXOntologyModelReference2021}.
Due to the open context of Automated Driving Systems, we assume that any domain ontology is incomplete.
Hence, to elaborate the concept presented in this paper qualitatively, we refrain from using a domain ontology such as the ASAM OpenXOntology.
In its assumption, the test case contract entails the operating conditions formalized as assertions and derive from the scenario description.

In turn, the test case's evaluation criterion aims to evaluate variables from the simulation run in order to generate a test result, for example, ``passed'' or ``failed''.
We propose to encode the validity requirement regarding the variables evaluated by the evaluation criterion in the test case contract's guarantee.
For example, a quantitative variable being used for the evaluation criterion may be associated with a quantitative uncertainty measure.
In this case, an upper bound for that uncertainty measure would be specified in the guarantee.
The purpose is to reflect that whenever the guarantee is satisfied a stakeholder can put confidence into the test result.
One approach to defining such a guarantee would be to ensure that the test result is unambiguous over the entire possible value range of the variables accepted by the guarantee.
How to obtain such formalizable guarantees that inspire credibility among stakeholders remains an open question for future work.

Furthermore, for future investigations, we propose using sensitivity analyzes of the test result with respect to the assumptions;
hence, supporting a quantitative statement about how strongly the test result's validity is influenced by the validity domains of the simulation models used.

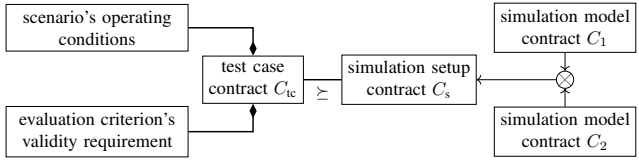
\begin{figure}[h]
    \centering
    \resizebox{\linewidth}{!}{
        \begin{tikzpicture}[
  cross/.style={path picture={
    \draw[gray]
    (path picture bounding box.south east) -- (path picture bounding box.north west) (path picture bounding box.south west) -- (path picture bounding box.north east);
}}
]

    \tikzstyle{taxonomy_rect} = [
      rectangle, 
      draw, 
      fill=white,
      align=center
    ];
  
    \tikzstyle{taxonomy_rect_ref} = [
      rectangle, 
      draw, 
      fill=lightgray, 
      fill opacity=0.8, 
      text opacity=1.0,
      minimum width=2cm, 
      minimum height=0.8cm, 
      align=center
    ];

    \tikzstyle{uml_composition} = [
      thick, 
      ->, 
      >=Diamond
      ];
    \tikzstyle{uml_aggregation} = [
      thick, 
      ->, 
      >={Diamond[open]}
      ];
    \tikzstyle{uml_inheritance} = [
      thick, 
      ->, 
      >={Triangle[open]}
      ];
    \tikzstyle{arrow} = [
      thick, 
      ->, 
      >=latex
      ];
    \tikzstyle{no_arrow} = [
      thick
    ];
  
    \node[taxonomy_rect, minimum width=3.5cm] (opcond) at (1,1) {scenario's operating \\ conditions};
    \node[taxonomy_rect, minimum width=3.5cm] (evalcrit) at (1,-1) {evaluation criterion's \\ validity  requirement};
    \node[taxonomy_rect] (ctc) at (4,0) {test case \\ contract $C_{\text{tc}}$};
    \node[taxonomy_rect] (cs) at (7,0) {simulation setup \\ contract $C_{\text{s}}$};
    \node[draw,circle,cross] (compose) at (10,0) {};
    \node[taxonomy_rect] (c1) at (10,1) {simulation model \\ contract $C_{1}$};
    \node[taxonomy_rect] (c2) at (10,-1) {simulation model \\ contract $C_{2}$};
  
    \draw[uml_composition] (opcond) -| (ctc);
    \draw[uml_composition] (evalcrit) -| (ctc);

    \draw[no_arrow] (ctc) -- (cs) node[midway, anchor=north, font=\small] {$\succeq$};
    \draw[->] (compose) -- (cs);
    \draw[->] (c1) -- (compose);
    \draw[->] (c2) -- (compose);

    \coordinate (top) at (0,2);
    \coordinate (bottom) at (0,-1.2);
    \path (top);
    \path (bottom);
  
  \end{tikzpicture}
    }
    \caption{Overview of the test case contract, simulation setup contract, and the composition of simulation model contracts.}
    \label{fig:summary_contracts_test_case_and_sim_model}
\end{figure}
\section{FRAMEWORK}
\label{sec:framework}

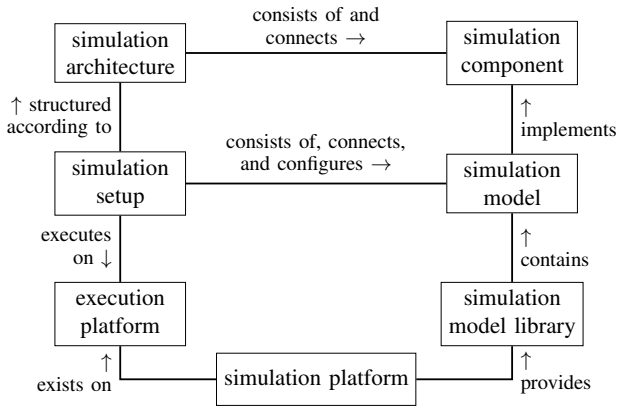
\begin{figure}[h]
    \centering
    \resizebox{0.95\linewidth}{!}{
        \begin{tikzpicture}

  \tikzstyle{taxonomy_rect} = [
    rectangle, 
    draw, 
    fill=white,
    minimum width=2cm, 
    minimum height=0.8cm, 
    align=center
  ];

  \tikzstyle{taxonomy_rect_ref} = [
    rectangle, 
    draw, 
    fill=lightgray, 
    fill opacity=0.8, 
    text opacity=1.0,
    minimum width=2cm, 
    minimum height=0.8cm, 
    align=center
  ];

  \tikzstyle{uml_composition} = [
    thick, 
    ->, 
    >=Diamond
    ];
  \tikzstyle{uml_aggregation} = [
    thick, 
    ->, 
    >={Diamond[open]}
    ];
  \tikzstyle{uml_inheritance} = [
    thick, 
    ->, 
    >={Triangle[open]}
    ];
  \tikzstyle{arrow} = [
    thick, 
    ->, 
    >=latex
    ];
  \tikzstyle{no_arrow} = [
    thick
  ];

  \node[taxonomy_rect] (simulation_architecture) at (-3,4) {simulation \\ architecture};
  \node[taxonomy_rect] (component) at (3,4) {simulation \\ component};
  \node[taxonomy_rect] (simulation_setup) at (-3,2) {simulation \\ setup};
  \node[taxonomy_rect] (simulation_model) at (3,2) {simulation \\ model};
  \node[taxonomy_rect] (execution_platform) at (-3,0) {execution \\ platform};
  \node[taxonomy_rect] (simulation_model_library) at (3,0) {simulation \\ model library};
  \node[taxonomy_rect] (simulation_platform) at (0,-1.0) {simulation platform};

  \draw[no_arrow] (simulation_platform) -| (execution_platform) node[pos=0.6, anchor=east, align=right, font=\small] {$\uparrow$ \\ exists on};
  \draw[no_arrow] (simulation_platform) -| (simulation_model_library) node[pos=0.6, anchor=west, align=left, font=\small] {$\uparrow$ \\ provides};
  \draw[no_arrow] (simulation_model) -- (simulation_model_library) node [midway, anchor=west, align=left, font=\small] {$\uparrow$ \\ contains};
  \draw[no_arrow] (simulation_model) -- (simulation_setup) node[midway, anchor=south, align=center, font=\small] {consists of, connects,\\ and configures $\rightarrow$};
  \draw[no_arrow] (component) -- (simulation_architecture) node[midway, anchor=south, align=center, font=\small] {consists of and \\ connects $\rightarrow$};

  \draw[no_arrow] (simulation_setup) -- (execution_platform) node[midway, anchor=east, align=right, font=\small] {executes\\ on $\downarrow$};
  \draw[no_arrow] (simulation_setup) -- (simulation_architecture) node[midway, anchor=east, align=right, font=\small] {$\uparrow$ structured \\ according to};
  \draw[no_arrow] (simulation_model) -- (component) node[midway, anchor=west, align=left, font=\small] {$\uparrow$ \\ implements};  

  \coordinate (top) at (0,5.0);
  \path (top);

\end{tikzpicture}
    }
    \caption{Context for and environment of the framework applying design contracts for simulation models and simulation setups, based on \cite{braunSupportingScenarioBasedValidationAutomated2022thesis}.}
    \label{fig:ontology_sim_arch_and_setup}
\end{figure}

Based on the contract definitions and operations defined in \cref{sec:contracts-for-simulation}, we propose a framework that uses these contracts to support the composition of simulation models into simulation setups.
The concepts relevant for this framework are described by the ontology in \Cref{fig:ontology_sim_arch_and_setup}.
In \cref{subsec:framework:composability_sim_arch,subsec:framework:composition_sim_setup}, we describe the configuration steps of the simulation setup, called the configuration phase.
We explain how computational resources can be considered during the configuration phase in \cref{subsec:framework:computational_resources}.
In \cref{subsec:framework:runtime_monitors}, we describe how functions can be generated to monitor whether the validity domains of the simulation models are satisfied during the simulation.

\subsection{Composability of the Simulation Architecture}
\label{subsec:framework:composability_sim_arch}

The simulation architecture defines the structure of components and their interconnects but does not necessarily preclude which simulation models to actually use, as several simulation models may be implementations of these components.
As the ports and contracts are defined for each component, the simulation architecture is checked for composability.
Composability is a prerequisite to parallel-compose the simulation models within the simulation setup.
The composability check ensures that the syntactic properties, such as data type and physical unit of the ports, match and that each variable has at most one source.
The composability check imposes a constraint on the available options for composing simulation models into a simulation setup.

\subsection{Composition of the Simulation Setup}
\label{subsec:framework:composition_sim_setup}

Compared to the simulation architecture, the simulation setup contains the actual simulation models that implement the components defined in the simulation architecture.
This important difference is pointed out in \Cref{fig:ontology_sim_arch_and_setup}.
A simulation model is associated with a contract for its component.
In addition, the simulation setup exposes the variables that are not controlled by any of the simulation models to, systemically speaking, the environment of the composition of simulation models.
As a result, these uncontrolled variables become inputs of the composition of simulation models and need to be parametrized externally.
In the context of scenario-based testing, the uncontrolled variables are parameters derived from the scenario.

The contracts associated with each component and their respective, syntactically suitable implementations (simulation models) are used to further constrain the available simulation models according to their validity domain.
The simulation model contracts represent the simulation models' validity domain, i.e., assumptions under which the simulation model is considered sufficiently valid and guarantees regarding validity metrics that represent the ``sufficient'' level of validity.
In return, a contract is derived from the test case.
This contract represents the scenario's operating conditions in the assumption and a validity requirement based on the evaluation criterion in the guarantee.
A knowledge base could be used to infer further facts from the scenario's description.
Analogously to the alphabet equalization step, see \cref{subsubsec:parallel-composition-and-quotient}, this step provides a semantic linkage between the scenario's description and the simulation models' contracts.
However, creating a consistent knowledge base that makes assumptions regarding the domain model transparent is a challenge on its own.
During the configuration phase of the simulation, i.e., during the composition of simulation models and their contracts, we consider simulation setups as sufficiently valid if the simulation setup contract refines the test case contract.

Let's consider a simple example where the simulation architecture consists only of two components, $I_1$ and $I_2$. $I_1$ is implemented by the simulation model $M_1$ which satisfies the contract $C_1$.
$I_2$ is implemented by two simulation models, $M_{2a}$ and $M_{2b}$, which satisfy the contracts $C_{2a}$ and $C_{2b}$, respectively.
The test case contract $C_{\text{tc}}$ entails the concrete scenario specification and the validity requirements toward the evaluation criterion, which is a variable controlled by the simulation setup.

One composable simulation setup, as illustrated in \Cref{fig:summary_contracts_test_case_and_sim_model}, would be the parallel composition of the simulation model contracts ${C_s = C_1 \otimes C_{2a}}$.
The parallel composition of contracts is used to evaluate the validity of a composition of simulation models, i.e., the simulation setup.
Whether the simulation setup is sufficiently valid depends on whether it refines the test case contract, i.e., whether ${C_s \preceq C_{\text{tc}}}$.
Given well-defined contracts, this step can be automated.
If the simulation setup is not sufficiently valid, the alternative composition ${C_s = C_1 \otimes C_{2b}}$ can be evaluated.

The example also illustrates how the quotient operator provides the inverse insight:
Based on a test case, the scenario entailed therein, and an existing residual simulation setup, the quotient operator can be used to derive the set of simulation models that are still missing in order to fulfill the test case's contract.
In other words, requirements for a to-be-developed simulation model can be derived based on the actual needs of that test case.
In the given example, the quotient operator can be used to derive the requirements for a simulation model $M_{2c}$ that implements a new contract ${C_{2c} \preceq C_{\text{tc}} / C_1}$.
That new contract represents requirements for the simulation model that are based on the test case under consideration.

\subsection{Accounting for Computational Resources}
\label{subsec:framework:computational_resources}

During the configuration phase of simulation setups, more than one sufficiently valid simulation setup may be identified.
It seems attractive to select the simulation setup that is the least computationally expensive.
We propose annotating the simulation models with a characterization of the computational resources required.
Selecting the least computationally expensive option may alleviate the scalability challenge of scenario-based testing.
For example, if two vehicle dynamics models, such as a bicycle model and a multi-body model, are available, the bicycle model may be computationally less expensive and hence be preferred, given that it is determined to be sufficiently valid for the test case at hand.

\subsection{Generation of (Runtime) Monitors}
\label{subsec:framework:runtime_monitors}

An important limitation of analyzing simulation models' validity domains during the configuration phase is the reliance on (1) well-defined contracts and (2) the assumption that the simulation models' behavior is indeed bounded by their validity domains.
On the first limitation, we recall that any modeling of the operational domain of Automated Driving Systems is necessarily incomplete.
As contracts are defined with reference to a model of the operational domain, i.e., physical quantities, entities, their relationships, and constraints, the contracts are necessarily incomplete as well.
Secondly, the assumption that the simulation model implements its contract dutifully may not always be valid.
Both implementation mistakes of the simulation model and unknown or unconsidered non-convexity of the validity metric may lead to an exit of the validity domain during runtime, i.e., the validity metric is not within the threshold specified in the contract guarantee.
To catch such violations of the validity domain during runtime, we propose to generate monitors based on the contracts which are defined for the simulation models.
These monitors can either be used during runtime or after the simulation -- in either case, they analyze the variables recorded during the actual simulation run.

Since the contracts defined for this paper are expressed as assertions, they can be used to generate monitors fairly easily.
During runtime or based on the variables recorded during the simulation run, the monitors substitute the variables from the recorded run into the assertions provided by the contracts.
If the assertions are violated, the runtime monitor can raise an alert and pinpoint to the variable and time step that raised the violation.
This automatic checking of contract violations during runtime may potentially reduce the need for expert knowledge to judge the validity of the simulation model after the execution of the test case.

\subsection{Discussion}
\label{subsec:discussion}

We expect that there will always be cases in which the inherent incompleteness of the domain model will lead to the generation of simulation setups that turn out to not be sufficiently valid at runtime.
This issue can be mitigated by generating monitors based on the contracts defined for the simulation models which, however, does not solve the underlying cause.
Furthermore, this approach requires the possibility to propagate the assertions and infer the validity domains through the simulation setup deterministically, which can become challenging for large simulation architectures.
In addition, capturing probabilistic properties of, for example, machine learning-based components, would require a different approach, such as probabilistic contracts.
Further work may also be needed to address co-simulation effects which are not yet captured in the current approach.
In the minimal example, the knowledge base acted as a ``bad bank'' for assumptions about and logical inference rules related to the operational domain.
However, even this step of making assumptions more transparent -- both on the scenario and domain model description side as well as on the simulation model side -- can be seen as helpful in making simulation results more credible to stakeholders.
\addtolength{\textheight}{-0.0cm} 

\section{CONCLUSION AND FUTURE WORK}
\label{sec:conclusion}

In scenario-based testing of Automated Driving Systems, a contract-based simulation setup framework has the potential to support the configuration of simulation setups that are sufficiently valid for the test cases they are intended to execute.
The concept capitalizes on the world-structuring character of scenarios for Automated Driving Systems and the structured description of simulation models' validity domains.
The presented approach might furthermore increase transparency regarding the assumptions made during the configuration of the simulation setup.
This transparency is important in order to justify trade-offs made, for example, between high-fidelity but expensive to compute, and low-fidelity but computationally efficient simulation models.
However, a new scalability issue arises with this approach, as well-defined contracts are required for each simulation model and test case.
Aside from the technical task, managing and implementing such contracts is an organizational challenge that requires both training and proper tool support.

Whether and how the concept can be made operational for productive simulation models and test cases remains an open question and will be addressed in future work.
In particular, an empirical validation of how well the approach supports handling complexity in simulation-based testing of Automated Driving Systems would be valuable.
Another interesting question is how the approach can be applied to large simulation model libraries and frameworks and how well it scales with the complexity of the domain model.
A logic inference method that is able to take conditional expressions into account could be a promising extension to the approach, especially for handling non-convex validity domains.
Another future direction is to investigate the application of probabilistic contracts and methods to propagate probabilistic properties through the simulation setup to support the generation of sufficiently valid simulation setups.

\section*{ACKNOWLEDGMENT}

We thank Mattias Nyberg (KTH Royal Institute of Technology) for the stimulating discussions.

\renewcommand*{\bibfont}{\footnotesize} 
\printbibliography

\end{document}
\typeout{get arXiv to do 4 passes: Label(s) may have changed. Rerun}